\documentclass[os, manuscript]{copernicus}

\usepackage{url} 
\usepackage[normalem]{ulem}

\usepackage{lineno}
\begin{document}

\nolinenumbers

\title{Sinking microplastics in the water column: simulations in the Mediterranean Sea}


\Author[1]{Rebeca}{de la Fuente}
\Author[1,2]{G\'abor}{Dr\'otos}
\Author[1]{Emilio}{Hern\'andez-Garc\'\i a}
\Author[1]{Crist\'obal}{L\'opez}
\Author[3,4]{Erik}{van Sebille}

\affil[1]{IFISC (CSIC-UIB), Campus Universitat de les Illes Balears, Palma de Mallorca, Spain}
\affil[2]{MTA--ELTE Theoretical Physics Research Group, Budapest, Hungary}
\affil[3]{Institute for Marine and Atmospheric Research, Utrecht University, Utrecht, the Netherlands}
\affil[4]{Centre for Complex Systems Studies, Utrecht University, Utrecht, the Netherlands}



\correspondence{Emilio Hern\'andez-Garc\'ia (emilio@ifisc.uib-csic.es)}

\runningtitle{Sinking microplastics in the water column}

\runningauthor{de la Fuente et al.}

\received{}
\pubdiscuss{} 
\revised{}
\accepted{}
\published{}


\firstpage{1}

\maketitle

\begin{abstract}
We study the vertical dispersion and distribution of negatively
buoyant rigid microplastics within a realistic circulation
model of the Mediterranean sea. We first propose an equation
describing their idealized dynamics. In that framework, we
evaluate the importance of some relevant physical effects:
inertia, Coriolis force, small-scale turbulence and
variable seawater density, and bound the relative error of
simplifying the dynamics to a constant sinking velocity added
to a large-scale velocity field. We then calculate the amount
and vertical distribution of microplastic particles on the
water column of the open ocean if their release from the sea
surface is continuous at rates compatible with observations in
the Mediterranean. The vertical distribution is found to be
almost uniform with depth for the majority of our
parameter range. Transient distributions from flash releases
reveal a non-Gaussian character of the dispersion and various
diffusion laws, both normal and anomalous. The origin of these
behaviors is explored in terms of horizontal and vertical flow
organization.
\end{abstract}

\copyrightstatement{To be inserted by journal staff}

\introduction  


Approximately 8 million tonnes of plastics end up in the oceans
every year \citep{jambeck2015plastic}. Nevertheless, only a
very small percentage, around $1 \%$, remains on the surface
\citep{van2015global,choy2019vertical}. The rest
leaves the surface of the ocean
\citep{ballent2013modeled,van2020physical}
through
beaching
\citep{turner2011occurrence}, biofouling
\citep{ye1991,chubarenko2016some,kooi2017ups} or sinking
\citep{erni2019distribution}, but also wind-driven mixing
presumably leads to
an underestimation for the amount of particles
remaining close to sea surface
\citep{kukulka2012effect,enders2015,suaria2016mediterranean,poulain2018small}.
The distribution of plastic pollution in the sea is poorly understood at present but would be
crucial to properly evaluate the exposure of marine biota to
this material, and formulate strategies for cleaning the oceans
\citep{horton2018microplastics}.



Floating plastics and those that have beached or sedimented on
the seafloor are relatively well studied through field
campaigns \citep[although explanation is missing for many findings;][]{andrady2017,erni2019distribution,kane2019}. In
contrast,
the presence of plastics within the water column has received
less attention, and many surveys in this realm are restricted
to so-called underway samples, a few meters below the surface
\citep[e.g.,][]{enders2015}. However, e.g.,
\citet{choy2019vertical} reported that below the mixed layer
and down to 1000~m depth in Monterey Bay, concentrations of
plastics are larger than at the surface
\citep{thompson2004lost,hidalgo2012microplastics}.
\citet{egger2020first} found more plastic between 5~m and
2000~m below the North Pacific Garbage Patch than at the
surface.
These findings turn out to mostly concern plastic pieces that,
according to their nominal material density, would be
classified as positively buoyant \citep{egger2020first}.


In this paper, we focus
on a certain class of plastic particles, negatively buoyant
rigid microplastics, excluding very small size,
and we estimate their vertical distribution
through the water column and their amount in the Mediterranean
Sea. Microplastic particles are among the most important
contributors to marine plastic pollution \citep{arthur2009}.
Closely following the work of \citet{monroy2017modeling} for
sinking biogenic particles but choosing particle properties to
correspond to those of negatively buoyant microplastics, we
first justify the use of a simplified equation of motion, in
which the plastic particle velocity is the sum of the ambient
flow velocity and a sinking velocity depending on particle and
water characteristics. In particular, we estimate the impact of
some corrections to this simple dynamics and evaluate in detail
the influence of the spatial variation of the seawater density
on the plastic dispersion and sinking characteristics. For our
Mediterranean case study, the impact of the varying seawater
density on particle trajectories can be comparable to the
estimated effect of the neglected small scales below the
hydrodynamical model's resolution.

We then estimate the amount
of microplastic particles
in the water column of the open Mediterranean. Our estimates
rely on a uniform vertical distribution, which is confirmed by our
numerical simulations to be a good approximation for
fast-sinking particles. This can be explained by a simple model
in which released particles sink with a constant velocity.
Detailed consideration of the transient dynamics identifies
small non-Gaussian vertical dispersion around this simple
sinking behavior, with transitions between anomalous and normal
effective diffusion.

\section{Types of microplastics in the water column}
\label{sec:classification}

The dynamics and the fate of microplastics in the ocean are
largely determined by their material density
\citep{erni2019distribution}. However, shape, size and rigidity
are also relevant properties, characteristic transport pathways
to the water column being different for different particle
types.

Typically, positively buoyant plastic types will remain
floating at the sea surface or close to it, and then will not
contribute to the microplastic content in the water column, the
topic we are interested in this paper. However, it has been documented experimentally that biofouling may increase sinking rates of particles up to 81\% and enhances sedimentation \citep{kaiser2017effects}. So, a class of high
abundance and mass may be represented by nearly neutrally
buoyant microplastic particles that are generated by biofouling
\citep{ye1991,chubarenko2016some} from positively buoyant
plastic types or by other mechanisms of aggregation with
organic matter, especially for small particle sizes \citep{kooi2017ups}.

In fact, the fallout from the North
Pacific Garbage Patch almost entirely consists of plastic types
nominally less dense than water \citep{egger2020first}.
Although some of these immersed particles finally reach the sea
bottom, their proportion in sedimented plastic is minor except
for the immediate vicinity of coasts where water is shallow.
Most of these particles remain in the photic zone
\citep{mountford2019eulerian,wichmann2019influence,soto2020hotspots}.
This suggests that reverse processes could also take place
after biofouling and that the dynamics of such particles is
complicated \citep{kooi2017ups,erni2019distribution}.

Particles denser than seawater dominantly accumulate at the sea
bottom \citep{mountford2019eulerian}. A mechanism by which
microplastics denser than water can also be present within the
water column is the finite time taken by them to reach the
bottom. Under continuous release at the surface and
sedimentation at the bottom, the transient falling would lead
to a steady distribution for the amount of plastic in the water
column at any given time, and this distribution has never been
estimated. Note that the Eulerian methodology of
\citet{mountford2019eulerian}, treating sedimentation (i.e.,
deposition on the seafloor) by parametrization and thus leaving
particles in the water column indefinitely long, is not
suitable for this estimation. One aim of this paper is to
explore this distribution by means of Lagrangian simulations.

There are different classes of microplastic particles denser
than seawater. For example,  dense synthetic
microfibers have been found to strongly dominate in sediment
samples far from the coast
\citep{woodall2014,brandt2015,bergmann2017,martin2017,peng2018},
and have been detected in large proportions in deep-water
samples and sediment traps in the open sea as well
\citep{bagaev2017,kanhai2018,peng2018,reineccius2020}. Mostly
originating from land-based sources
\citep{dris2016,carr2017,gago2018,wright2020}, it is not
obvious to explain their abundance on abyssal oceanic plains
\citep{kane2019}. Maritime-activity sources \citep{gago2018}
can contribute to that. Another reason could be that their
special and deformable shape results in a strongly reduced
settling velocity \citep{bagaev2017} that allows long distance
horizontal transport \citep{nooteboom2020}. In any case,
it is difficult to estimate the amount of microfibers in the oceans
due to sampling issues and to their absence from statistics of
mismanaged plastic waste \citep{carr2017,barrows2018}, and we
will not consider them further in this paper. We also disregard
 films, which are only sporadically
encountered in the open ocean \citep{bagaev2017} and thus have moderate
importance.

We concentrate in the following on dense rigid
microplastic particles. The most abundant particles of this
class are fragments \citep[e.g.,][]{martin2017,peng2018}, which
have an irregular shape, but their extension is usually
comparable in the three dimensions.
Experimental estimates for the settling velocities of irregular
fragments or other nonspherical particles have suggested
considerable deviations from values predicted by the Stokes law
\citep{kowalski2016,khatmullina2017,kaiser2019}, so that it is
unclear how a precise full equation of motion should be
constructed. For a qualitative exploration of particle
transport through the water column, we will argue in
Sect.~\ref{subsec:corrections} and App.~\ref{sec:APPnonspherical} that the Maxey--Riley--Gatignol (MRG)
equation \citep{maxey1983equation} should be appropriate for a
reasonably wide range of such particles.


Whatever their precise equation of motion is, these sinking
particles (directly detected by \citet{bagaev2017} and
\citet{peng2018}) are thought to reach the seafloor relatively
fast \citep{chubarenko2016some,kane2019,soto2020hotspots},
landing within horizontal distances of tens of kilometers from
their surface location of release (see
Sect. \ref{subsec:salinity} and
App. \ref{sec:APPeffects}). One consequence of their fast sinking
is the absence of almost any fragmentation after they
leave the sea surface \citep{andrady2015,corcoran2015}, and the
influence of biological processes on the particles' properties
should also be moderate, leaving their size and
shape intact during sinking. Note that, in contrast to the case of floating plastics \citep{kooi2017ups,kvale2020global}, interaction of sinking plastics with particulate matter of biological origin appears to be moderate. This is according to the absence of a need to disassemble microplastic pieces from biological aggregates during sample processing as described by \citet{bagaev2017}. Note, however, that experimental results by \citet{michels2018} indicate that aggregation with organic material might occur within a sufficiently short time at surface layers, which would likely lead to increased sinking velocities \citep{long2015interactions}.
Transport by bottom currents
\citep{kane2019,kane2020seafloor} is important for explaining
their distribution in sediments after coastal release. However,
the statements above imply that the dense rigid
microplastic content of samples from deep-sea trenches, abyssal
plains \citep{cauwenberghe2013,brandt2015,peng2018,kane2019}
must
originate from sources at the surface of the open sea rather
than from coastal inputs.

While methodological issues make the quantification of
abundance difficult
\citep{song2014,andrady2015,filella2015,lindeque2020},
negatively buoyant microplastic fragments have indeed been
found in surface and near-surface samples of the open waters of
the Mediterranean Sea \citep{suaria2016mediterranean} and the
Atlantic Ocean \citep{enders2015}, respectively, from which
they can contribute to microplastic content of the water column and
deep-sea sediments \citep{brandt2015,bagaev2017}. Horizontal
transport of these particles can be carried out by marine
organisms, and spontaneous attachment to pieces of positive
buoyancy is a further possibility but is not yet discussed in
the literature.
Composite pieces of debris or those that contain trapped air
(including foams in some cases) may also represent a source
of microplastic ending at the water column \citep{andrady2015}. However, most of such particles are
presumably released by local maritime activity. An example of
this are flakes of paint and structural material from boats and
ships, which contain negatively buoyant alkyds and
poly-(acrylate/styrene). Despite the particle's high density,
large amounts of them may be found in the sea surface microlayer
where surface tension keeps them floating \citep{song2014}.
The range of horizontal transport of these particles at the sea
surface is unclear, but expected to be restricted to short
distances because sinking from the sea surface microlayer is
considerable, especially in waters disturbed by waves
\citep{hardy1982,stolle2010}.


While the idea of \citet{kooi2019} to treat all plastic
particles together by means of continuous distributions is
appealing, the above considerations strongly favor the
separate treatment of positively buoyant pieces, negatively
buoyant microfibers, and negatively buoyant rigid particles
of sufficiently big size,
since these classes have very different dynamics and sources.
In the following we concentrate on the properties, amount and
dynamics of
particles of the last class.

\section{Considerations for modeling negatively buoyant rigid microplastics}

\subsection{Physical properties}
\label{subsec:physical}

\begin{figure}[t]
\includegraphics[width=8.3cm]{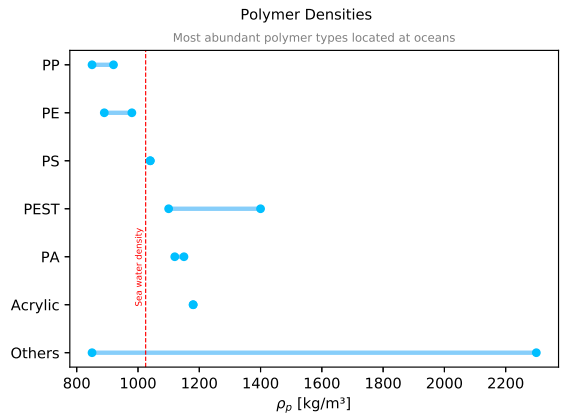}
\caption{\label{Fig:clas}Polymer densities for the most abundant microplastics identified in water bodies \citep{erni2019distribution}.}
\end{figure}

From a meta-analysis of 39 previous studies,
\citet{erni2019distribution}
established the proportion of the most abundant polymer types
discharged into water bodies: PE (polyethylene, 23 \%), PP
(polypropylene, 13 \%), PS (polystyrene, 4 \%) and PP\&A (group
of polymer types formed by polyesters, PEST, polyamide, PA and
acrylics, 13 \%). Note that these proportions do not
distinguish between different regions (e.g., coastal region or
open water; even inland water bodies of urban environments are
included in the analysis) and between the particle types (size
range and shape) concerned in the different studies. We
organize these polymer types
according
to their density
\citep{chubarenko2016some,andrady2017,erni2019distribution} in Fig.
\ref{Fig:clas}: PP between $850-920 \ kg/m^3$, PE $890-980 \
kg/m^3$, PS with $1040 \ kg/m^3$ (excluding its foamed version), PEST in the range $1100-1400
\ kg/m^3$, PA within $1120-1150 \ kg/m^3$ and acrylic with
$1180 \ kg/m^3$. There is also some less abundant plastic like
polytetrafluoroethylene (PTFE) which has higher densities, in
the range $2100-2300 \ kg/m^3$.

Thus, the full range of microplastic particle densities in the
ocean, denoted here as $\rho_{p}$, is $850-2300 \ kg /m^3$, and
most of them have densities within the interval $850-1400 \
kg/m^3$. This has to be compared with the seawater density,
which close to the surface has a conventional mean value of
$\rho_{f}=1025 \ km/m^{3}$ (red line in Fig. \ref{Fig:clas})
and changes around $1 \% $ from the surface to the sea bottom.
Since we are interested in sinking material, and for the sake
of maximal practicality, we restrict our
study to microplastics of densities
$1025 \ kg /m^3 < \rho_{f} < 1400 \ kg /m^3$.


Another relevant property of plastic particles is their
size. By a widely accepted definition, microplastics are particles with a
diameter less than $5~mm$ without any lower limit \citep{arthur2009}.
Some observations at the ocean surface
show that the most common diameter is around $1~mm$
\citep{cozar2014,cozar2015plastic}, with an exponential
decay with increasing diameter up to $100~mm$. However, the
absence of this peak in other studies that show an increasing
abundance with further decreasing size
\citep{enders2015,suaria2016mediterranean,erni2017} suggests
\citep{song2014,andrady2015,erni2017,bond2018,lindeque2020}
the  need for new technologies in sampling methods (which
usually use trawl nets with a mesh size around $0.3~mm$) and
especially for the adaptation of careful and standardized
analysis procedures to avoid artifacts \citep{filella2015}.


Field data about distributions of size and
quantifiers of shape for negatively buoyant rigid particles in
the water column or deep-sea sediments are not available to
date to the best of our knowledge, except in the Artic for
\citet{bergmann2017}. However, their results may not apply to
the majority of the oceans because of the very special dynamics
provided by melting and freezing of sea ice
\citep{bergmann2017}. Data
from \citet{bergmann2017} and
\citet{song2014} about unspecified sedimented fragments and
paint particles, respectively, exhibit an increasing abundance
with decreasing size, most particles being smaller than
$0.05~mm$. Laboratory findings about surface degradation of individual particles
also indicate such a tendency \citep{song2017}. Thus, these findings
seem to indicate the prominent presence of small pieces of plastic.
Nevertheless the observations of Bagaev et al (2017), Kanhai et al.
(2018) and Peng et al. (2018) do not indicate this abundance of small
particles.

For these reasons, we will disregard particles of extremely small size.
To keep our qualitative study sufficiently simple, we will consider all
our modelled particles to have a
radius $a = 0.05~mm$ (a diameter of $0.1~mm$). This is a rather small
size, but still within the commonly measured ranges. As we will discuss
in Section~\ref{subsec:corrections}, this radius is
well within the validity range of the MRG equation.

\subsection{Source estimation}
\label{subsec:source}

In this subsection we indicate the total amount of
dense microplastics entering
 the water column in open waters of the Mediterranean. Despite the correlation of plastic source with coastal population density, the rapid fragmentation of small particles along the shoreline \citep{pedrotti2016changes} and the seasonal variability of spatial distribution of floating particles \citep{macias2019surface}, we focus on local maritime activity and exclude direct
release from surface accumulation areas or the coast, either from urban areas or from
rivers. The estimations are based on the
results of \citet{kaandorp2020closing}.
They provide a total amount of yearly plastic release into the Mediterranean
in the range $2200-4000$ tonnes, from which
around 37\% corresponds to negatively
buoyant plastic, and 6\% are due to maritime activity.
 This 37\% agrees well with
previous global estimations \citep{lebreton2018evidence}.

We will take these numbers, 4000 tonnes per year,
37\% of sinking particles,
 and the proportion of direct release by maritime
activity (6\%) to obtain in Sect. \ref{subsec:verticaldist} an
estimate for the basin-wide yearly release of negatively
buoyant sphere-like microplastics in the open Mediterranean.
Note that we choose the upper bound, 4000 tonnes per year, in order to
account for the considerable amount of unregistered particles.

\subsection{Dynamics}
\label{subsec:dynamics}

A standard modeling approach
\citep{siegel1997trajectories,monroy2017modeling,liu2018influence,monroy2019}
for the transport of noninteracting sinking particles
is to consider the time-dependent particle
velocity $\vec{v}$ as the combination of the ambient
fluid flow $\vec{u}$ and a
settling velocity $\vec{v}_{s}$ as:
\begin{equation}
\vec{v}= \vec{u} + \vec{v}_{s} \ ,
\label{eq:reference}
\end{equation}
with
\begin{equation} \vec{v}_{s}=(1-\beta)\vec{g} \tau_{p} \ , \
\beta=\frac{3\rho_{f}}{2\rho_{p}+\rho_{f}} \textrm{
 , and  } \tau_{p}=\frac{a^{2}}{3 \beta \nu} \ .
\label{eq:vsbeta}
\end{equation}
$\vec{g}$ denotes the gravitational acceleration vector,
pointing downwards; $\beta$ is a parameter depending on the
particle and the fluid densities, $\rho_p$ and $\rho_f$,
respectively. Particles heavier than water have $\beta<1$, and
$\beta=1$ for neutrally buoyant particles. The expression given
for $\beta$ assumes spherical particles. $\tau_{p}$ is the
Stokes time, i.e., the characteristic response time of the
particle to changes in the flow, where $a$ is the radius of the
particle and $\nu$ the kinematic viscosity of the fluid.
Although Eq. (\ref{eq:reference}) is commonly used, we are not
aware of a systematic justification of it in the microplastics
context. This will be done in Section \ref{subsec:corrections}.

\subsection{Numerical procedures}
\label{subsec:numerical}

For the flow velocity $\vec{u}$ we use a 3D velocity field from
NEMO (Nucleus for European Modelling of the Ocean), which
implements a horizontal resolution of 1/12 degrees and 75
s-levels in the vertical with updates data every 5 days
\citep{madec2008nemo,madec1996global}. Salinity and
temperature are also extracted from that model. The Parcels
Lagrangian framework \citep{delandmeter2019parcels} is used to
integrate the particle trajectories from Eq.
(\ref{eq:reference}) or more complex ones to be considered in
Sect. \ref{subsec:corrections}. Typical numerical experiments
to obtain the results presented below consist of distributing a
large number $N$ of particles in a horizontal layer over the
whole Mediterranean on
the nodes of a sinusoidal-projection grid
\citep{seong2002sinusoidal}, so that their release is with uniform horizontal density.
We locate this input source at
1 m depth to avoid surface boundary conditions. After particles
are released at some initial date, in a so-called flash
release, they evolve under equations of motion such as Eq.
(\ref{eq:reference}), and the statistics of the resulting
particle cloud are analyzed.

\section{Results}
\label{sec:results}

\subsection{Range of validity of Eq. (\ref{eq:reference})}
\label{subsec:corrections}

We next show, closely following the treatment of
\citet{monroy2017modeling} for biogenic particles, that
possible inertial effects that would correct Eq.
(\ref{eq:reference}) are negligible for the sizes and
densities of typical dense microplastics. To this end,
similarly to many other studies
\citep{michaelides2003hydrodynamic,balkovsky2001intermittent,cartwright2010dynamics,haller2008inertial},
we start by choosing the simplified standard form of the more
fundamental Maxey--Riley--Gatignol (MRG) equation
\citep{maxey1983equation}, and analyze under which
conditions it is valid for microplastic transport. After
finding the MRG equation to be valid for an important range of
microplastic particles, we will explore its relationship with
Eq. (\ref{eq:reference}).

The simplified MRG equation gives the velocity $\vec{v}(t)$ of
a very small spherical particle in the presence of an external flow $\vec{u}(t)$ as
\begin{equation}
\frac{d\vec{v}}{dt}=\beta \frac{D\vec{u}}{Dt}+ \frac{\vec{u}-\vec{v}+\vec{v}_{s}}{\tau_{p}}.
\label{eq:MRG}
\end{equation}
Beyond sphericity, two conditions are needed for the validity of Eq.
(\ref{eq:MRG})
\citep{monroy2017modeling,maxey1983equation}: a) the
particle radius, $a$, has to be much smaller than the Kolmogorov
length scale $\eta$ of the flow, which has values in the range
$0.3 \ mm < \eta < 2 \ mm $ for wind-driven turbulence in the
upper ocean
\citep{jimenez1997ocean}; b) the particle Reynolds number
$Re_{p}= \frac{a| \vec{v }-\vec{u}|}{\nu} \approx \frac{a
v_s}{\nu}$ should satisfy $Re_p \ll 1$. Note that this last
condition imposes restrictions on the values of the particles'
density and size, partially via the settling velocity $v_s=|\vec{v}_{s}|$.
For the most abundant sinking microplastics, i.e., with densities
$\rho_{p} = 1025-1400 \ kg/m^3$, we now determine the range of
validity of Eq. (\ref{eq:MRG}) assuming $\nu = 1.15 \times 10^{-6} \
m^{2}/s$ and $\rho_{f} = 1025 \ kg/m^3$ to be fixed. This gives
$\beta$ in the range $0.8-1$. The possibility of small changes
in the seawater density as the particle sinks, which translates
to variations in $\vec{v}_s$, will also be analyzed in Section
\ref{subsec:salinity}.

\begin{figure}[t]
\includegraphics[width=8.3cm]{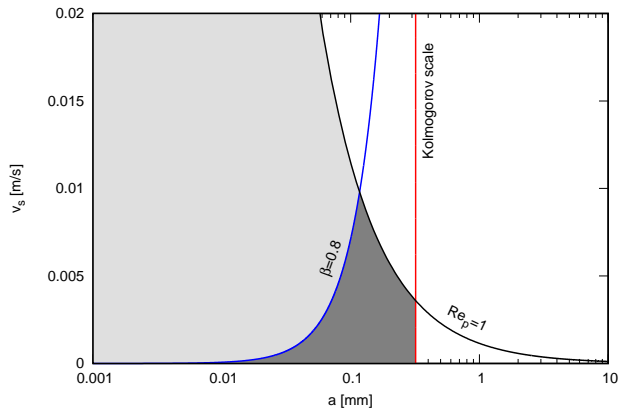}
\caption{\label{Fig:diagrama}Settling velocities and particle sizes for which
Eq. (\ref{eq:MRG}) holds. Kolmogorov scale is represented by the
red line and $Re_p = 1$ with a black line, which bound the area
of validity. Blue curve corresponds to $v_s = v_s (\beta, a)$
for $\beta = 0.8$, the upper bound to $v_s$ for typical microplastic
densities. Dark shading denotes the plastic particle sizes and
corresponding settling velocities for which application of Eq.
(\ref{eq:MRG}) is valid.}
\end{figure}

In Fig. \ref{Fig:diagrama} we show a diagram with the settling
velocities and particle sizes for which Eq. (\ref{eq:MRG}) is
valid. We plot the minimal value of the Kolmogorov scale
$\eta=0.3 \ mm$ with the red line \citep{jimenez1997ocean}, and
$Re_p = 1$ with a black line, which bound the area of validity
(shaded in the plot). We also indicate $v_s$ as a function of
$a$ for $\beta = 0.8$ with the blue curve, corresponding to the
upper bound to $v_s$ for typical microplastic densities. In
total, the zone with \emph{soft} shading in Fig.
\ref{Fig:diagrama} represents a parameter region where Eq.
(\ref{eq:MRG}) applies for particles with $\beta < 0.8$ (i.e.,
particles falling faster than the typical ones), whereas the
area of our interest, corresponding to $\beta \geq 0.8$, is
represented by a \emph{dark} shading, denoting the typical
plastic sizes and corresponding settling velocities for which
the equation is valid. As a rule of thumb, in a typical
situation, validity of Eq. (\ref{eq:MRG}) requires $v_{s} <
0.01 \ m/s$ and $a < 0.3 \ mm$. As discussed in
Section~\ref{subsec:physical}, information about particles in
the validity range is particularly sparse for surface waters
because of the usual sampling techniques, but sediment data
indicates the prevalence of sufficiently small particles.
Furthermore, in sufficiently calm waters, the Kolmogorov scale
is larger (of the order of millimeters,
\citet{jimenez1997ocean}), so that $a$ can be increased to this
size without compromising the equation validity. These
estimates of the Kolmogorov scale anyway assume wind-driven
turbulence and are thus restricted to the mixed layer
\citep{jimenez1997ocean}, below which $\eta$ is undoubtedly
larger.
Deviations from a spherical shape may lead to a more complicated motion than that described by the MRG equation, especially through particle rotation \citep{voth2017}. In App.~\ref{sec:APPnonspherical}, we present quantitative arguments for the applicability of the MRG equation to rigid microplastic particles of common shapes in the parameter ranges of our interest.


The simplified MRG equation, Eq.
(\ref{eq:MRG}) thus represents an appropriate basis for qualitative
estimations of the transport properties of negatively buoyant rigid
microplastics in the water column.
Note that rigidity of the particles is an essential condition
which is why the advection of microfibers is out of the scope of
this paper.


The connection between Eq. (\ref{eq:MRG}) and its approximation
Eq. (\ref{eq:reference}) is made by noticing that $\tau_{\eta}
\approx 1 s$ in the ocean
\citep{monroy2017modeling,jimenez1997ocean}, so that the Stokes number
$St = \tau_p /\tau_{\eta}$, which measures the importance of
particle inertia in a turbulent flow, is very small (of the
order of $10^{-3}-10^{-2}$). Thus an expansion of the MRG
equation for small $St$ (smallness of the Froude number, i.e.
smallness of fluid accelerations with respect to gravity, is
also required) can be performed. The expansion in its simplest form leads
to \citep{balachandar2010,monroy2017modeling,drotos2019}:
\begin{equation}
\vec{v} \approx \vec{u} + \vec{v}_{s} + \tau_{p}(\beta-1)\frac{D\vec{u}}{Dt}.
\label{eq:inertial}
\end{equation}

We can now take the results of \citet{monroy2017modeling} for
biogenic particles of sizes and densities similar to the
microplastics considered here to show that the inertial
corrections (the term proportional to $\tau_p$) in Eq.
(\ref{eq:inertial}) are negligible, so that the simpler Eq.
(\ref{eq:reference}) correctly describes sinking of
microplastics in the considered parameter range. For
completeness, we report in App. \ref{sec:APPeffects} the explicit
numerical calculations showing this (in which the influence of
the Coriolis force is also taken into account, since it is
known to be of the same order or larger than the inertial term
when a large-scale flow is used for $\vec{u}$). In particular
we find from release experiments from $1\ m$ below the surface
of a large number of particles with $\beta$ in the range
$0.8-1$ in the whole Mediterranean that the
difference between horizontal particle positions after
10 days of integration calculated from Eq. (\ref{eq:inertial})
and the simpler Eq. (\ref{eq:reference}) is just a 0.26\% of
the horizontal displacements. For the vertical motions the
difference is of about 0.05\%. Thus, Eq. (\ref{eq:reference})
provides a proper description of the dynamics.

Even if an equation of motion is accurate, the accuracy of its
solution is limited by that of the input data. In particular,
small-scale flow features are absent from oceanic velocity
fields $\vec{u}$ simulated on large-scale domains, which is an
important limitation of the respective solutions of Eq.
(\ref{eq:reference}). The NEMO velocity field of our choice is
not an exception, but a rigorous evaluation of the
corresponding errors of particle trajectories is not possible
without knowing the actual small-scale flow. Nevertheless, one
can roughly estimate the effect of these small scales by adding
a stochastic term to Eq. (\ref{eq:reference}) with statistical
properties similar to the expected ones for a small-scale flow
\citep{monroy2017modeling,kaandorp2020closing}. Results similar
to those by \citet{monroy2017modeling}, summarized in App.
\ref{sec:APPeffects}, indicate that after 10 days of
integration the relative difference between particle positions
given by Eq. (\ref{eq:inertial}) with and without this `noise'
term modeling small scales (using $\beta=0.99$) is around 8\%
for the horizontal displacements and 5\% for the vertical ones.
The figures become 12\% and 5\%, respectively, when evolving
the particles for 20 days. These errors are moderate, although
they may be of importance under some circumstances
\citep{nooteboom2020}. We consider these figures as a baseline
to evaluate corrections to the simple Eq. (\ref{eq:reference}):
adding more complex particle-dynamics terms to it will not
improve plastic-sedimentation modeling unless the effect of
these corrections is significantly larger than the above
estimations for the effect of the unknown small-scale flow. In
the following we consider the simple Eq. (\ref{eq:reference}),
but we estimate the implications of assuming or not a constant
value of the water density.

\subsection{Effect of variable seawater density}
\label{subsec:salinity}

In this section we analyze the role of a variable seawater
density on the particle settling dynamics. Fluid density is
calculated from the TEOS-10 equations, which is a thermodynamically
consistent description of seawater properties derived from a
Gibbs function, for which absolute salinity is used to describe
salinity of seawater and conservative temperature replaces
potential temperature
\citep{pawlowicz2010every}. In the simulations described in this section, as
particles move in the ocean they encounter different
temperatures and salinities, as given by the NEMO model
described in Sect. \ref{subsec:numerical}, and then they
experience different values of the ambient-fluid density.

We consider particles
of a fixed density $\rho_{p}=1041.5 \ kg/m^3$. This implies
that for a nominal water density of $\rho_{f}= 1025 \ kg/m^3$
the value of $\beta$ would be $\beta=0.99$, giving a sinking
velocity $v_s=6.2~m/day$ for our particles of radius $a = 0.05~mm$, but this sinking velocity will be
increased or decreased in places where water density is lower
or higher, respectively, so that we have a spatially- and
temporally-dependent velocity in Eq. (\ref{eq:reference}). The
particle density and size have been chosen to be representative
of the slowly-sinking microplastic particles, for which we
expect the seawater density variations to have the largest
impact. In this way we find some upper bound for the importance
of variability in seawater density for particle trajectories.

\begin{figure*}[t]
\includegraphics[width=18cm]{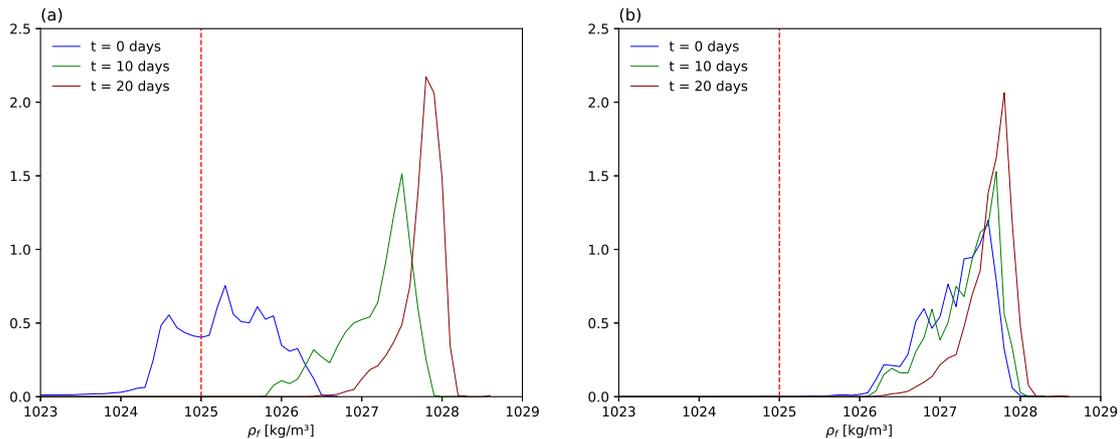}
\caption{\label{Fig:beta}Normalized histogram of the seawater density $\rho_f$ at the positions of $N=78803$ particles after $t= 0$, $10$, and $20$ days of being released over the whole Mediterranean. (a): summer release (release date 8 July 2000). (b): winter release (release date 8 January 2000). The particles' density is fixed at $\rho_p=1041.5 kg/m^3$, and fluid density is obtained from the TEOS-10 equations. The vertical line indicates a conventional seawater density of $1025 kg/m^3$.}
\end{figure*}

We release $N=78,803$ particles
over the whole Mediterranean Sea,
and monitor their trajectories under
Eq. (\ref{eq:reference}). The left panel of Fig. \ref{Fig:beta}
shows the histogram of water densities encountered by the
particles when the release is performed on July 8th 2000. On this
summer date, the Mediterranean is well stratified, at least in
its upper layers. Initially the particles are in surface waters
with a range of salinities that average approximately to the
nominal $\rho_{f}= 1025 \ kg/m^3$. But as they sink in time
they reach layers with higher densities (and more homogeneous
across the Mediterranean). When the release is done in winter
(right panel of Fig. \ref{Fig:beta}) the water column is more
mixed, so that the range of water densities experienced by the
particles released at different points is always narrow. But
the mean water density turns out to be always larger than the
conventional surface density of $\rho_{f}= 1025 \ kg/m^3$, so
that a slightly slower sinking is expected to occur.


\begin{figure*}[t]
\includegraphics[width=18cm]{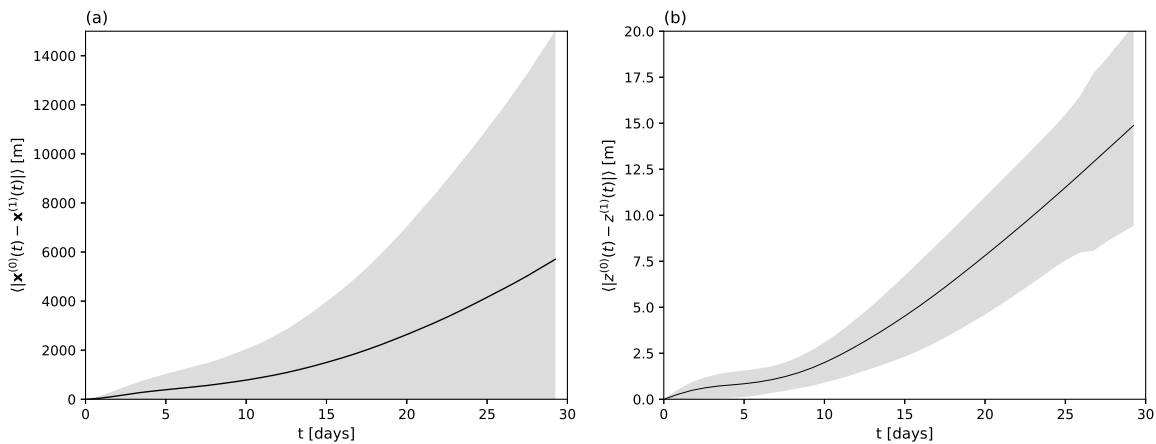}
\caption{\label{Fig:densitydependent}The distance, as a function of time, between trajectories obtained with
constant nominal fluid density of $\rho_f=1025 kg/m^3$ and the actual variable fluid
density, both starting at the same initial location. The range of the values among all
particles released in different points of the Mediterranean is indicated by the shaded
area, while the solid line indicates the average over the particles. Particles have
$\rho_{p}=1041.5 kg/m^3$, and all parameters are the same as for the summer release
in Fig. \ref{Fig:beta}.(a): horizontal distances; (b): vertical distances.}
\end{figure*}

We illustrate the impact of this variable density on particle
trajectories for the summer release in Fig
\ref{Fig:densitydependent}. Here we compute, as a function of
time, the range of horizontal  $|\textbf{x}^{(0)}- \textbf{x}^{(1)}|$ and vertical $|z^{(0)}-z^{(1)}|$ distances and its
average among particles. Trajectories $x^{(0)}(t)$ and
$x^{(1)}(t)$ are obtained with constant nominal fluid density
($1025~kg/m^3$) and position-dependent fluid density,
respectively, using the same release location and date 8 July
2000 in both cases. Particle density is fixed at
$\rho_{p}=1041.5 \ kg/m^3$. The difference between the two
calculations (and thus the error of considering that constant
value for the density) should be compared to average horizontal
and vertical displacements of $95 \ km$ and $124 \ m$,
respectively, at $t = 20$ days. At that time, we thus find that
the influence of variable fluid density on the dynamics is
about $3 \%$ for the horizontal movement and $6 \%$ for the
vertical displacement on average.

%

\begin{table*}[t]
\caption{\label{tab:densityeffect} Relative effect on horizontal and vertical particle positions after 10 and 20 days of integration, averaged over 78803 particles released over the whole Mediterranean at $1~m$ depth, of replacing the actual seawater density by a nominal value $\rho_f=1025~kg/m^3$.}
\begin{tabular}{|ll|c|c|}
\tophline
                                &             & 10 days & 20 days   \\
\middlehline
\multirow{2}{*}{Summer release} & Horizontal: & 1.12 \%                     & 2.75 \%                       \\
                                & Vertical:   & 3.19 \%                     & 6.25 \%                       \\
\middlehline
\multirow{2}{*}{Winter release} & Horizontal: & 1.88 \%                     & 5.62 \%                       \\
                                & Vertical:   & 8.14 \%                     & 9.32 \%                       \\
\bottomhline
\end{tabular}
\end{table*}

A summary of the average relative differences on horizontal and
vertical particle positions between using the
location-dependent seawater density and a nominal constant
value $\rho_f=1025 kg/m^3$, both in winter and summer periods,
is displayed in Table
\ref{tab:densityeffect}. The relative error produced by
assuming a constant density is larger in the vertical
direction. It is also larger for the release in winter, but
this is a consequence of taking a value for the reference density
that is not representative of winter waters but is strongly biased (see Fig.
\ref{Fig:beta}, right). If using a reference value more
appropriate for winter waters (say $\rho_f\approx 1027~kg/m^3$)
the relative error remains quite small, due to the weaker
stratification of the sea during this season.
In fact, the reference value is also biased in the summer unless the investigation is restricted to the surface.

In brief, we
see that the effect of location-dependent density may be a
relevant effect to evaluate microplastic transport.
At least, the traditional value of seawater density may be biased, which may be reflected in the particle trajectories.
We recall,
however, that we used parameters for the particle properties
for which they are slowly falling. The impact of variable
density on particles that sink faster will be smaller. Also, the
effects reported in Table \ref{tab:densityeffect} remain of the
order of the estimations of the effects of unresolved small
scales of the flow (Sect. \ref{subsec:corrections}). As a
consequence, in the following we will not consider variable
seawater density, but restrict our modeling to Eq.
(\ref{eq:reference}) with a constant nominal value of the
sinking velocity $v_s$.


\subsection{Total mass and vertical distribution of microplastics}
\label{subsec:verticaldist}


We will first estimate the total mass of negatively buoyant
rigid microplastics in the water column of the open
Mediterranean Sea by assuming a uniform vertical distribution,
then we will justify this assumption by running numerical
simulations according to the conclusion of
Section~\ref{subsec:corrections} about the equation of motion.

For estimating the total mass, we take the quantities of
Section~\ref{subsec:source} ($4000~tonnes/year$ of plastic
release, with 37\% being negatively buoyant of which 6\%
originates from maritime activities) to compute the rate $r$ at
which microplastic particles of our interest enter the water
column in the open sea: $r=4000~tonnes/year \times 0.37 \times
0.06 = 88.8~tonnes/year$, or $r=0.24~tonnes/day$.

The next step is to estimate the time during which these microplastic
particles remain in the water column before reaching the sea
bottom. We take the mean depth for the Mediterranean to be
$h=1480~m$ \citep{ETOPO1data,ioc2008bodc} and estimate a residence time $\tau$ as the
time of sinking to that mean depth.
The residence time depends on the sinking velocity,
$\tau=h/v_s$, and thus on the physical properties of the
microplastic particles. Assuming a seawater density
$\rho_f=1025~kg/m^3$, and the range of plastic densities and
their proportions described in Sect. \ref{subsec:physical}, we
see from Eq. (\ref{eq:vsbeta}) that for microplastic particles
of radius $a\approx 0.05~mm$ the range of sinking velocities is
$6.20 - 509.23~m/day$, giving a residence time in the range
$3.1 - 255~days$. Averaging these times weighted by the
proportion of each type of plastic we get $\overline
\tau\approx 14~days$. Combining the input rate $r$
with this mean residence time we get an estimate for the total
amount
present in the water column at any given time as
$Q=r\overline\tau$: the result is $Q\approx 3.36$ tonnes of
dense rigid microplastics if all of them would be
in the form of particles of size $a=0.05~mm$. This is below but
close to 1\% of the estimated upper bound of $470$ tonnes of
floating plastic in the Mediterranean (according to the
corresponding estimation of \citet{kaandorp2020closing}).

We  emphasize the many uncertainties affecting this result
(Sections~\ref{subsec:physical} and \ref{subsec:source}), and
we highlight the one related to particle size: because of the
quadratic dependence of the sinking velocity on the particle
radius $a$, Eq. (\ref{eq:vsbeta}), choosing the particle size
to be half of the one used here will lead to a four times
larger estimate for the mass if the same release rate is
assumed. This enhanced retention of smaller particles in the
water column may imply, depending on the actual size
distribution, a dominance of very small particles on the
plastic mass content of the water column.
 However, our estimates of plastic input
into the ocean (we use mainly \citet{kaandorp2020closing}) rely
on observations that do not catch extremely small particles.
These considerations
further justify our choice of a radius
$a=0.05~mm$, small but still easily detectable, as convenient
to provide reasonable estimations of negatively buoyant rigid
microplastic mass in the water column within commonly quoted
size ranges. We can not exclude larger plastic content at
smaller sizes. Another source of bias may be not considering in this study the impact of small-scale turbulence and convective mixing events. While small-scale turbulence might cause an increase of lifetimes of particles in the water column, dense water formation
and rapid convection, a process reported in areas such as the Gulf of Lions, might likely reduce particle retention time. These events take place in winter and were shown to transfer particles from the ocean surface to mid-waters
(1000 meters) and deep ocean (>2000 meters) in a very short time (1-2 days) and lead to the formation of bottom
nepheloid layers \citep{de1999slope,vidal2009across,heussner2006spatial,stabholz2013impact}.


The result for the total mass is independent of the horizontal
distribution of particle release, which is quite inhomogeneous \citep[Fig.~1 of][]{liubartseva2018tracking}. However, for a
rough estimate of the density of these microplastics in the
water column,
we assume a uniform particle distribution over the whole
Mediterranean both in horizontal and in vertical. Since the
volume of the Mediterranean is about $4.39\times 10^6~km^3$
\citep{ETOPO1data} the estimated density
would be $\rho_V\approx 7.7\times 10^{-11}~kg/m^3$
(with the above-discussed scaling issues with $a$).
We remind the reader that
this is a value for the open sea,
and our study does not address coastal areas, where the density
would likely be higher.

The above estimates are rather rough
as a result of the mentioned uncertainties.
The assumption of a uniform distribution in the vertical
direction has not yet been justified either,
but we will show it to be appropriate by means of
our
simulations of particle release starting at $1~m$ depth over
the whole Mediterranean. Instead of performing a continuous
release of new particles at each time step, and computing
statistics over this growing number of sinking particles, we
approximate this by the statistics of all positions at all time
steps of a set of particles deployed in a single release event.
This assumes a time-independent fluid flow, but this approximation
is appropriate, since the dispersion of an ensemble of particles
released in a single event follows rather well-defined
statistical laws, see Section~\ref{subsec:transient}, and is
thus independent of the time-varying details of the flow.
Particles are removed when touching the bottom. For our
estimate, we use $\beta = 0.8$, i.e., assuming the fastest
sinking velocity of typical plastic particles, for which particles reach vertical depths deeper if compared with the slower sinking velocities used in our study.

\begin{figure}[t]
\includegraphics[width=8.3cm]{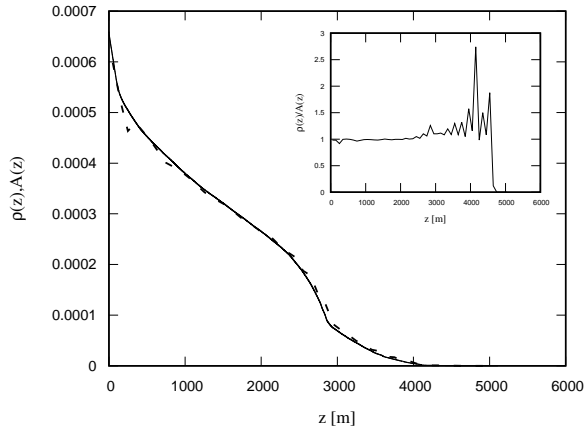}
\caption{\label{Fig:densityz}The continuous line is the area that the Mediterranean
has at each depth $z$. The dashed line is
microplastic density per unit of depth $\rho(z)$ under
continuous release of particles with $\beta=0.8$ at $1~m$ depth. Both curves have
been normalized to have unit area, so that they can be compared on
the same scale. The binning size is $100~m$. The inset shows the ratio $\rho(z)/A(z)$,
proportional to the mass density of microplastic per unit of volume $\rho_V(z)$.}
\end{figure}

Figure \ref{Fig:densityz} shows $\rho(z)$, the density of
plastic particles per unit of depth $z$ in the whole
Mediterranean, and also $A(z)$, the amount of area that the
Mediterranean has at each depth $z$. Both functions have been normalized such that the value of their integrals with respect to $z$ is one; in this way the functions can be displayed in the same plot. We see that both curves are nearly identical (in
fact, just proportional, because of the normalization),
indicating that the variation of the number of particles with
depth is essentially due to the decrease of sea area with
depth. A clearer way to see that is to plot $\rho(z)/A(z)$,
which is proportional to the mean plastic concentration per
unit volume at each depth $z$, $\rho_V(z)$. We see that this
quantity is nearly constant, at least in the first 3000~m. At
larger depths a weak increase seems to occur, but made unclear
by the poor statistics arising from the small area and number
of particles present at these depths. Thus, the hypothesis of a
uniform distribution of plastic in the water column seems to be
a reasonable description of the simulation of the
fastest-sinking particles.

A uniform distribution of plastics in $z$ is what is
expected if the vertical velocity of the particles is exactly a
constant (since each particle will spend exactly the same time at
each depth interval). The equation of motion used, Eq.
(\ref{eq:reference}) corrects this constant sinking velocity
$\vec{v_s}$ with a contribution $\vec{u}$ from the ambient flow.
Thus, the close-to-constant character of the plastic concentration
may imply that the flow correction $\vec{u}$ is negligible, at
least when considering its effect over the whole Mediterranean.
Another possibility is that the fluctuating
flow component $\vec{u}$ in Eq. (\ref{eq:reference}) results in a vertical dispersion
compatible with a constant concentration. Although the former explanation
predicts an alteration from a constant if the settling velocity is
sufficiently small to allow $\vec{u}$ to induce a stronger vertical dispersion,
we will see in the next section that a nearly constant concentration may be
assumed for the majority of our parameter range.

\subsection{Transient evolution}
\label{subsec:transient}


We now analyze in detail the
transient evolution of particle
clouds initialized by flash
 releases at a fixed depth.
Numerically we proceed by
releasing $N=78803$ particles
uniformly distributed over the
entire Mediterranean
 surface at $1\ m$ depth in the winter
season, as already described. They evolve
according to Eq. (\ref{eq:reference}) using a constant water
density. We take
three examples for the particle density, which correspond to
$\beta = 0.8$, $0.9$, $0.99$, or $v_s=153.48$, $68.21$, $6.20$
m$/$day for our particles of radius $a = 0.05~mm$, respectively; in what follows, these setups shall be
denominated as v153, v68 and v6. The horizontal displacements
during the particle sinking times are much larger (of the order
of $60~km$) than the sea depth, so that in fact the particles
are sinking \emph{sideways} \citep{siegel1997trajectories}.
However, the horizontal displacements still remain very small
compared to the basin size, and we concentrate on the vertical
motion. Even though the vertical steady distribution has been
found to be close to uniform in
Section~\ref{subsec:verticaldist}, the reason for this is not
evident,
and we will give support here for the pertinence of
this finding to most of the relevant parameter range.

\begin{figure}[t]
\includegraphics[width=8.3cm]{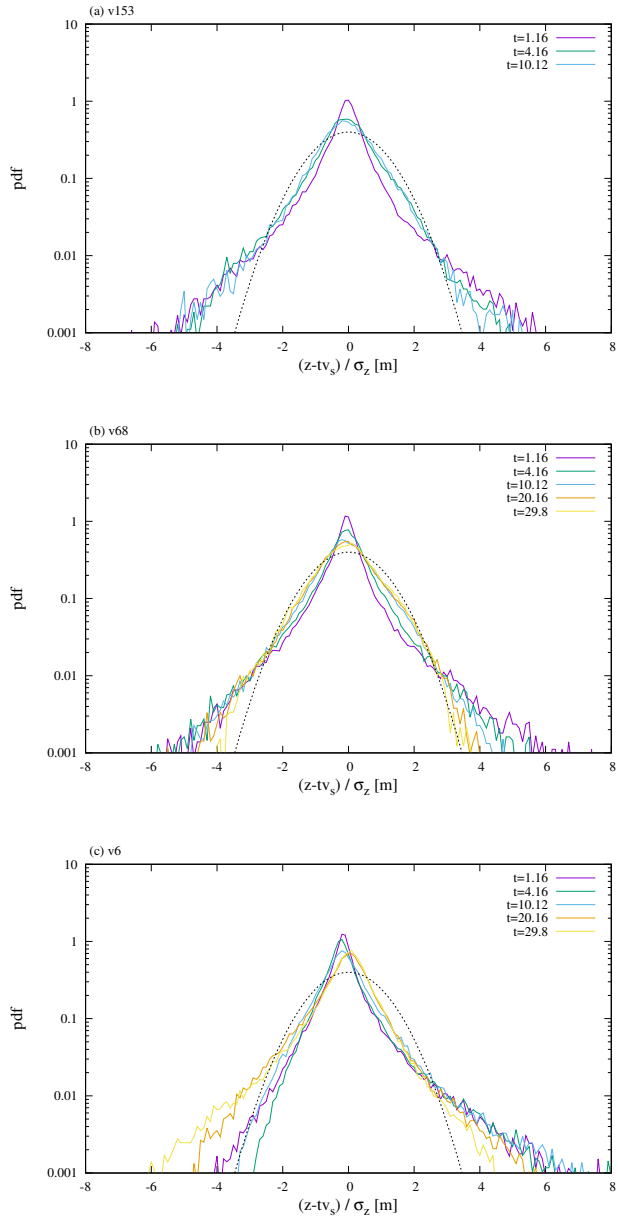}
\caption{\label{Fig:histograms}The probability density function, estimated from a histogram of bin size $0.1$,
of all particles released in the Mediterranean in the rescaled variable
$\tilde{z}=\frac{z-tv_s}{\sigma_z}$ for the different setups (v153, v68 and v6) and times (in days) as indicated. For comparison, normal distributions of zero mean and unit variance are shown with a dashed line.}
\end{figure}

Figure \ref{Fig:histograms} shows the vertical particle
distribution at different times (upper plot is for v153, middle
for v68 and bottom for v6). The plot is given in terms of a
rescaled variable $\tilde{z}=\frac{z-tv_s}{\sigma_z}$ where
$\sigma^2_z \equiv \langle ( z_i - \langle z_i \rangle )^2
\rangle$ is the variance of the particles' $z$ coordinate. Here
the subindex $i$ refers to the particle and $\langle \ldots
\rangle$ denotes averaging over different particles. Thus, we
plot in the figure the rescaled distribution of the particles
around the average depth of the particles at any given time.
For comparison, the normal distribution is plotted with dashed
lines. This figure shows deviations from Gaussianity for early
times. The deviation from normal distribution decreases for later instants
but remains considerable,
especially for the tails, which may also be indicative of anomalous
diffusive behavior. For reference, particles reach the mean
Mediterranean depth, $h=1480~m$ at times $\tau=9.64$, $21.8$
and $246.7$ days for v153, v68 and v6, respectively.



Since a non-Gaussian distribution is usually linked to
anomalous dispersion \citep{neufeld2009}, we now analyze this aspect in detail by
considering how the variance of the vertical particle
distribution, $\sigma^2_z (t)$, evolves.
Although there is a continual loss of particles because of
reaching the seafloor with a varied topography, we illustrate
in App. \ref{sec:APPbathymetry} that our conclusions are likely
unaffected by this effect.

\begin{figure}[t]
\includegraphics[width=8.3cm]{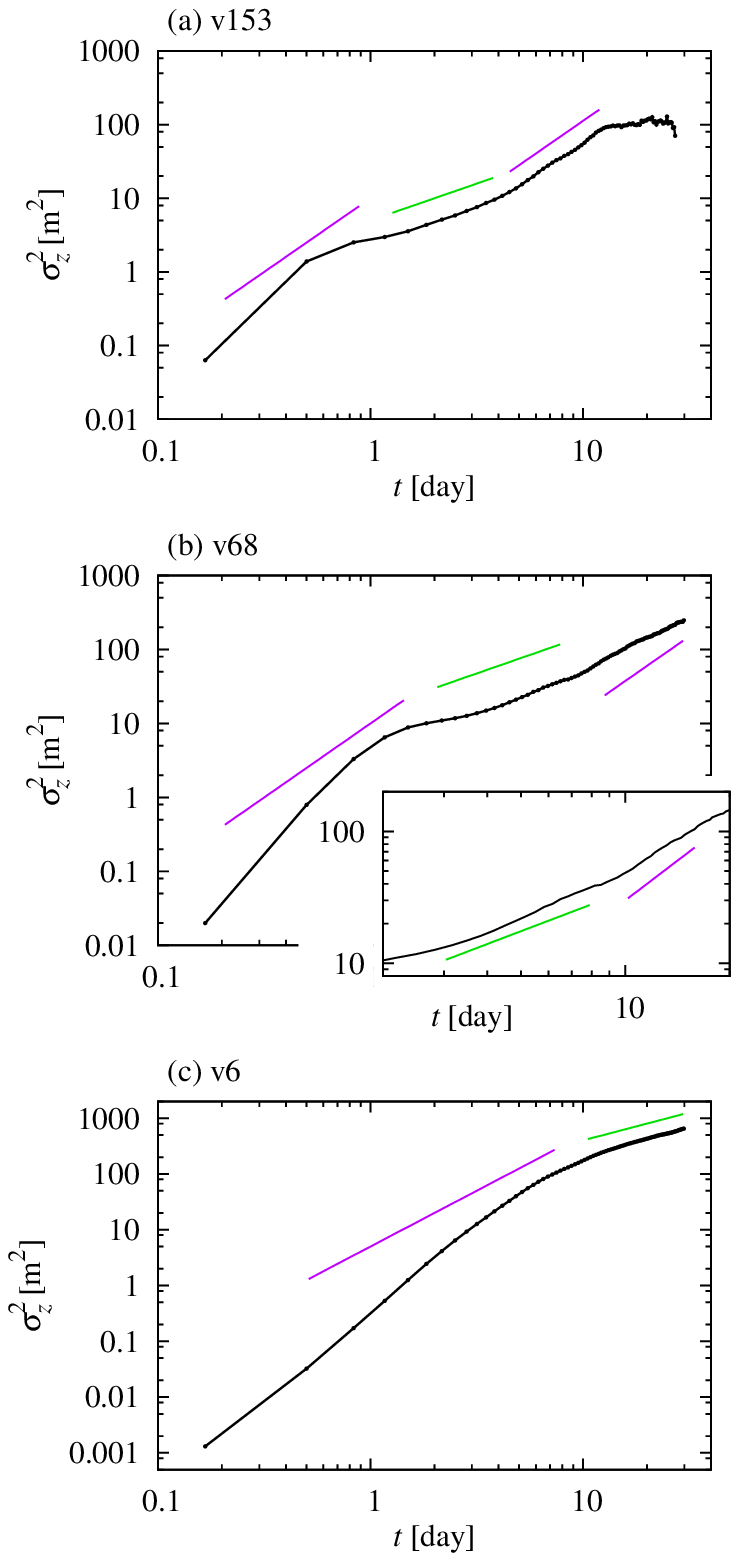}
\caption{\label{Fig:variancetime}Variance of depth reached by the particles as a function of time.
Straight lines represent power laws for reference, with exponents $1$ (in green, corresponding to standard diffusion)
and $2$ (in purple, corresponding to ballistic dispersion).}
\end{figure}

According to Fig. \ref{Fig:variancetime}, dispersion appears to
be governed by different laws in different regimes, which we
shall distinguish by the approximate effective exponents $\nu$,
defined through approximate behaviors $\sigma^2_z \sim t^\nu$
in different time intervals.


We start our analysis with the fastest-sinking particles (v153,
Fig.~\ref{Fig:variancetime}a). At the very beginning,
superdiffusion takes place with $\nu > 2$, which may be related
to autocorrelation in the flow, but we will iterate on this
question when comparing different settling velocities. Around
$t = 1$ day, the evolution seems to become consistent with
normal diffusion ($\nu = 1$), usual after initial transients in
oceanic turbulence
\citep{berloff2002material,reynolds2002}. However, around
$t = 4.5$ days, we can observe a crossover to ballistic
dispersion ($\nu = 2$).

We explain this last crossover as resulting from a different
mean sinking velocity in diverse regions of the Mediterranean,
associated with up- and down-welling. This can be modeled in an
effective way by writing the vertical position of particle $i$
as
\begin{equation}
z_i = \langle z_i \rangle + \bar{\omega}_i t + W_i ,
\label{eq:decomposition}
\end{equation}
where $\langle \ldots \rangle$ denotes, as before, an averaging
over different particles. Here we are assuming that $z_i -
\langle z_i \rangle$ evolves according to the sum of a constant
average \emph{velocity} contribution $\bar{\omega}_i$ for
sufficiently long times (a characteristic of the flow region
traversed by particle $i$), and of $W_i$, a Wiener process
representing fluctuations with zero mean and defining a
diffusion coefficient $D_i$ for each trajectory by
$\overline{W_i^2} = D_i t$. The overbar refers to temporal
averaging for asymptotically long times along the trajectory of
a given particle (but assuming that the particle remains in a
region with a well-defined $\bar{\omega}_i \neq 0$), and $D_i$
characterizes the strength of the fluctuations.
Assuming $\langle \bar{\omega}_i W_i \rangle = 0$,
\begin{equation}
\sigma^2_z \equiv \langle ( z_i - \langle z_i \rangle )^2 \rangle = \langle \bar{\omega}_i^2 \rangle t^2 + \langle D_i \rangle t ,
\end{equation}
that is, the variance is a sum of a ballistic and a normal
diffusive term, associated with regional differences in the
mean velocity and with fluctuations, and dominating for long
and short times, respectively. Writing $D = \langle D_i
\rangle$, the crossover between the two regimes is obtained by
equating the two terms as
\begin{equation}
t^* = \frac{ D }{ \langle \bar{\omega}_i^2 \rangle } .
\label{eq:crossover}
\end{equation}

\begin{figure}[t]
\includegraphics[width=8.3cm]{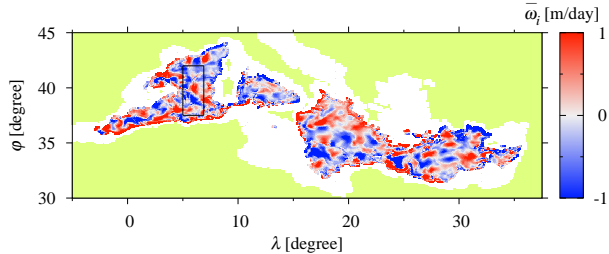}
\caption{\label{Fig:08HorizontalVm}$\bar{\omega}_i$ as estimated from $t = 10.83$ days
plotted at the initial position of each particle $i$ in the v153 simulation.
The black rectangle in the Western Mediterranean is the area of large depth considered
in App. \ref{sec:APPbathymetry}.}
\end{figure}

\begin{figure}[t]
\includegraphics[width=8.3cm]{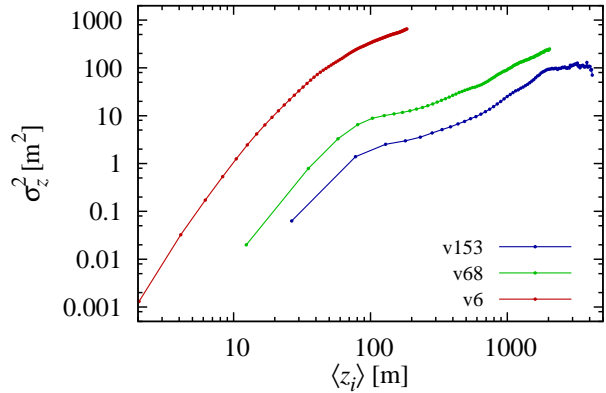}
\caption{\label{Fig:variancedepth}Variance of depth reached by the particles as a function of their mean depth.}
\end{figure}

To evaluate Eq. (\ref{eq:crossover}), we first estimate
$\bar{\omega}_i$ for each particle from the ``asymptotically''
long time of $t = 10.83$ days, which is the latest time after
the crossover still in the ballistic regime in Fig.
\ref{Fig:variancetime}a, when the contribution of fluctuations
should already have become negligible. The horizontal pattern of the
estimated $\bar{\omega}_i$ is presented in Fig.
\ref{Fig:08HorizontalVm}, which confirms its patchiness
throughout the Mediterranean, associated with mesoscale features.
Computing $\langle
\bar{\omega}_i^2 \rangle$ and fitting a line to $\sigma^2_z(t)$
between $t = 1.4$ and $4$ days to estimate $D$, we obtain $t^*
\approx 4.5$ days from Eq. (\ref{eq:crossover}), which
remarkably agrees with Fig. \ref{Fig:variancetime}a. After
approximately $t=12$ days there is hardly any dispersion, since
most of the particles are close to the sea bottom (cf.
Fig.~\ref{Fig:variancedepth}) where the vertical fluid velocity is
nearly zero. Note also a small drop in $\sigma^2_z($ at the very end of the time
series, where the results may actually be subject to artifacts,
see App. \ref{sec:APPbathymetry}. However, this is of
minor importance, since the distribution of particles so close
to the bottom should anyway be strongly influenced by
resuspension and remixing by bottom currents
\citep{kane2020seafloor}.

The different regimes are not as clear in the v68 case as for
v153, see Fig. \ref{Fig:variancetime}b.
One evident novel feature is a subdiffusive regime
during the transient from the initial superdiffusion (as in the
case of horizontal tracer dispersion in the ocean studied by
\citet{berloff2002material,reynolds2002}).
Approximate normal diffusion is then observed until $t = 10$ days,
when a crossover to a faster dispersion does seem to take place, see the inset.
A fit of normal diffusion from
$t = 4$ to $8$ days and the velocity variance at $t = 12.5$
days give an estimate $t^* \approx 11.7$.
However, the long-time ballistic regime is not clear. In fact, a long-term return from such a ballistic regime to
normal diffusion is expected as a result of increasing
horizontal mixing, which renders $\bar{\omega}_i$ time
dependent and makes it approach zero.
According to a careful visual inspection of the inset in
Fig.~\ref{Fig:variancetime}b, this may take place already
around $t = 14$ days.

For v6 (Fig. \ref{Fig:variancetime}c), the transition from the
initial superdiffusive regime to that of normal diffusion
appears to not involve subdiffusion.
This is already informative: the fluid velocity field is the
same for the three simulations with different settling
velocity, so the differences must originate from the different
rate of sampling of the different fluid layers by particles
while they sink. In particular, the decay of autocorrelation is
obviously faster for faster-sinking particles, since it is
determined by the spatial structure of the velocity field.

While this is one possible explanation for the earlier timing
of the initial transition from anomalous to normal diffusion for
higher settling velocity, one cannot exclude that a
depth-dependent organization of the flow is more in play; note
that $\nu > 2$ at the beginning, which might not be explained
by simple autocorrelation but might be characteristic of
properties of the velocity field at those depths. The governing
role of the spatial structure is supported by Fig.
\ref{Fig:variancedepth}: the transition in question takes place
at the same depth ($\approx 100$m) in the different
simulations, which seems to point to mixed-layer processes.
Depth-dependence might also govern the suppression of ballistic
dispersion for long times, but it is very unclear.

Note in Fig. \ref{Fig:variancedepth} that the vertical variance
is not expected to grow much larger for v68 than for v153 even
if the simulation were longer. Therefore, even if the constancy
of the steady vertical distribution relies on the weak vertical
dispersion for v153 (see Section \ref{subsec:verticaldist}),
constancy is expected to hold in most of our parameter range. A
considerably stronger dispersion and a possible corresponding
deviation from constancy may arise only for extremely low
settling velocities, like for v6 in Fig.
\ref{Fig:variancedepth}.




\conclusions  

We have discussed the different types of plastics occurring in
the water column, pointing out gaps in our knowledge about the
sources, transport pathways and properties of such particles.
It would be highly
beneficial to have distributions of size, polymer type and
quantifiers of shape recorded separately for the dynamically
different classes of microplastics.

We have focused our attention on rigid microplastic particles
with negative buoyancy.
We have argued that the simplified MRG
equation approximates the dynamics of such particles
sufficiently well for qualitative estimations.

We have then analyzed the importance of different effects in
this equation, and concluded that the Coriolis and the inertial
terms are negligible. When a velocity field of large-scale
nature is input to the equation (such that small-scale
turbulence is not resolved), or when the variability in
seawater density is neglected, moderate but possibly
non-negligible errors emerge \citep{nooteboom2020}.
However, our conclusions about the vertical distribution and
dispersion of microplastics rely on robust features of the
large-scale flow and must remain unaffected by moderate errors.
We also note that the traditional value of seawater density,
$\rho_{f}=1025 \ kg/m^{3}$, is representative only for
near-surface layers in the summer, and correcting for the bias
could  reduce the error of simulations with a constant
seawater density. A suitable equation of
motion for the particles considered is constructed by adding to
the the external velocity field a constant settling term, as
also found by \citet{monroy2017modeling} for marine biogenic
particles.

When the velocity field of the Mediterranean sea is
approximated by realistic simulation, this equation of motion
results in a nearly uniform steady distribution along the water
column, perhaps except at extremely low settling velocities.
The corresponding total amount of plastic present in the water
column is relatively small, close to 1\% of the floating plastic
mass, but it may be an important contribution to the
microplastic pollution in deep layers of the ocean, and is
subject to several uncertainties.

Note that only those microplastic particles are considered here
that have not yet sedimented on the bottom, and the plastic
amount sedimented on the seafloor is  large
\citep{brandt2015,liubartseva2018tracking,peng2018,mountford2019eulerian,soto2020hotspots}.
The suitability of our equation of motion to describe the
sinking of a class of microplastic particles implies that
advection by the flow may contribute to large-scale horizontal
inhomogeneity of deep-sea plastic sediments by means of
recently described noninertial mechanisms
\citep{drotos2019,monroy2019,sozza2020accumulated}. This
may especially be so in regions where redistribution by bottom
flows is restricted to small distances, like abyssal plains
\citep{kane2019}.
Resuspension and
redistribution may be dominant in forming sedimented patterns
\citep{kane2020seafloor}, and a future investigation should
take all processes into account to
identify zones of high plastic concentration on the sea bottom.

As for the vertical distribution profile, its approximate
uniformity may be linked to the weak vertical dispersion of
particles that is found in our simulations started with a flash
release over the whole surface of the Mediterranean sea. The
shape of the emerging transient vertical distribution exhibits
deviations from a Gaussian, which are related to anomalous
diffusive laws that dominate the vertical dispersion process in
some phases.

The different diffusive laws are related to the properties of
the decay in the Lagrangian velocity autocorrelation defined
along the trajectories of the sinking particles. An important
example is the transition from initial superdiffusion to a
longer phase of normal diffusion, occurring around $100$ m
depth, which indicates that the particles enter into a
different flow regime. Another characteristic of the velocity
field is a horizontal patchiness, which results in a long-term
ballistic dispersion as long as the particles' horizontal
displacements remain small. The vertical diffusion returns to
the normal type when horizontal mixing becomes more developed.
These results suggest regional
differences in the sinking process, so that regional modeling might be more
appropriate than a whole-basin approach.
Future studies will include different areas of the oceans, and
analyze the role of Lagrangian coherent structures on the
different vertical dispersion regimes.



\dataavailability{The velocity field from the NEMO simulation used in our study can be downloaded from \url{http://opendap4gws.jasmin.ac.uk/thredds/nemo/root/nemo_scan_catalog.html}. Parcels is a set of Python classes developed under the TOPIOS project and is accessible at \url{https://oceanparcels.org/}. The scripts for running the particle transport simulations are available upon request from Rebeca de la Fuente, \url{rebeca@ifisc.uib-csic.es}.} 




\appendix


\section{Deviations from a spherical particle shape}
\label{sec:APPnonspherical}

We quantitatively assess the impact of deviations from a spherical shape through a correction to the settling velocity $v_\mathrm{s}$. The simplified MRG equation, Eq.~(\ref{eq:MRG}), or its first-order approximations in the Stokes number, Eqs.~(\ref{eq:inertial}) and (\ref{eq:inertialC}), are affected by particle geometry through the drag force and the added mass term; however, accelerations are irrelevant for $v_\mathrm{s}$, so that the added mass term does not appear in its formulation or in the simple approximation of Eq.~(\ref{eq:reference}). We will compare values of the settling velocity describing nonspherical and spherical particles with the same density, then finally comment on the results' relevance for Eqs.~(\ref{eq:MRG}), (\ref{eq:inertial}) and (\ref{eq:inertialC}).

Most generally, the settling velocity vector $\mathbf{v}_\mathrm{s}$ can be obtained by balancing the drag force $\mathbf{F}_\text{drag}(\mathbf{v}-\mathbf{u})$ (a function of the difference of the particle and the fluid velocities, $\mathbf{v}$ and $\mathbf{u}$, respectively) with the resultant of gravitational and buoyancy forces:
\begin{equation}\label{eq:v_s}
	0 = \mathbf{F}_\text{drag}(\mathbf{v}-\mathbf{u}) + V \left( \rho_\mathrm{p} - \rho_\mathrm{f} \right) \mathbf{g}
\end{equation}
with $\mathbf{v}-\mathbf{u}=\mathbf{v}_\mathrm{s}$, where $V$ is the particle's volume, $\rho_\mathrm{p}$ and $\rho_\mathrm{f}$ are the densities of the particle and the fluid, respectively, and $\mathbf{g}$ is the gravitational acceleration vector. For a spherical particle with radius $a$, the Stokes drag force reads as
\begin{equation}\label{eq:drag_sph}
	\mathbf{F}^\text{(sph)}_\text{drag}(\mathbf{v}-\mathbf{u}) = - 6 \pi \mu (\mathbf{v}-\mathbf{u}) a ,
\end{equation}
where $\mu$ is the dynamical viscosity of the fluid. According to \citet{leith1987,ganser1993}, an appropriate approximation for small nonspherical particles is
\begin{equation}\label{eq:drag_non}
	\mathbf{F}^\text{(non)}_\text{drag}(\mathbf{v}-\mathbf{u}) = - 6 \pi \mu (\mathbf{v}-\mathbf{u}) \left( \frac{1}{3} a_\mathrm{n} + \frac{2}{3} a_\mathrm{s} \right) ,
\end{equation}
where $a_\mathrm{n}$ is the radius of the sphere with equivalent area projected on the plane perpendicular to the relative velocity $\mathbf{v}-\mathbf{u}$, and $a_\mathrm{s}$ is the radius of the sphere with equivalent total surface. From either of the last two equations, the settling velocity is obtained by substituting $\mathbf{v}-\mathbf{u} = \mathbf{v}_\mathrm{s}$, and solving Eq.~(\ref{eq:v_s}) for $\mathbf{v}_\mathrm{s}$. We denote the magnitudes of the settling velocities obtained from Eq.~(\ref{eq:drag_sph}) and Eq.~(\ref{eq:drag_non}) by $v_\mathrm{s}^\text{(sph)}$ and $v_\mathrm{s}^\text{(non)}$, respectively.

To characterize the correction in the settling velocity for a given nonspherical particle (with a given density $\rho_\mathrm{p}$) with respect to assuming a spherical shape with a radius $a$, we will consider
\begin{equation}\label{eq:q}
	q \equiv \frac{v_\mathrm{s}^\text{(non)}}{v_\mathrm{s}^\text{(sph)}} = \frac{3}{4\pi} \frac{V^\text{(non)}}{a^2 \left( \frac{1}{3} a_\mathrm{n} + \frac{2}{3} a_\mathrm{s} \right)} ,
\end{equation}
where $V^\text{(non)}$ is the real volume of the given particle. In order to evaluate Eq.~\eqref{eq:q}, one has to specify the shape and the size of the particle, its orientation with respect to its relative velocity, and also how $a$ is derived from its real size.

Note that it is always possible to define an $a$ for which $q = 1$, i.e., for which there is no correction arising from the deviation from a spherical shape. In this sense, any choice of $a$ representing a spherical shape, including ours in the manuscript, describes the settling velocity of certain nonspherical particles, the question is just their shape and size, which will mutually depend on each other for a given $a$. We will nevertheless proceed by choosing a shape class and defining $a$ along independent considerations, because we intend to link a given $a$ to a single particle size as identified during the processing of field observations.

The shape of rigid microplastic particles is not usually described in the literature, but we can see photographs of some examples in, e.g., \citet{song2014,brandt2015,bagaev2017}. For an explorative computation, a reasonable choice seems to be a rectangular cuboid with edges $A$, $B = \hat{B} A < A$ and $C = \hat{C} A < B < A$, where one or both of $\hat{B}$ and $\hat{C}$ are less than $1$ but greater than, say, $0.1$.

Under this assumption, the particle size will correspond to the longest edge, $A$, of the cuboid if the size is identified through microscopy as the largest extension \citep[``length''; e.g.,][]{cozar2015plastic}; and it may be related more to the middle edge, $B$, if one thinks of a sieving technique \citep[e.g.,][]{suaria2016mediterranean}. The naive choice will be $a = A/2$ and $a = B/2$ in these two cases.

We can substitute either of these choices of $a$ in Eq.~\eqref{eq:q}, as well as the appropriate formulae describing the actual cuboid. $V = ABC$ is unique, and so is $a_\mathrm{s}$,
\begin{equation}\label{eq:r_s}
	a_\mathrm{s} = \left( \frac{AB+AC+BC}{2\pi} \right)^\frac{1}{2} .
\end{equation}
However, $a_\mathrm{n}$ depends on the particle's orientation with respect to the relative velocity. Implications will be discussed when interpreting the results, and we take here all three directions parallel to edges $A$, $B$ and $C$ to represent different possibilities. The corresponding expressions for $a_\mathrm{n}$ read as
\begin{equation}\label{eq:r_n}
	a_{\mathrm{n},X} = \left( \frac{ABC}{\pi X} \right)^\frac{1}{2}
\end{equation}
for $X = A$, $B$ and $C$.

After substituting all these expressions in Eq.~\eqref{eq:q}, we obtain
\begin{align}
	q_X^{(A/2)} &= 9\pi^{-\frac{1}{2}} \hat{B}\hat{C} \left[ \left(\frac{\hat{B}\hat{C}}{\hat{X}}\right)^\frac{1}{2}+2^\frac{1}{2}\left(\hat{B}+\hat{C}+\hat{B}\hat{C}\right)^\frac{1}{2} \right]^{-1} , \\
	q_X^{(B/2)} &= \hat{B}^{-2} q_X^{(A/2)} ,
\end{align}
where $X = \hat{X} A$ has been introduced.

\begin{figure}[h!]
	\centering
	\includegraphics{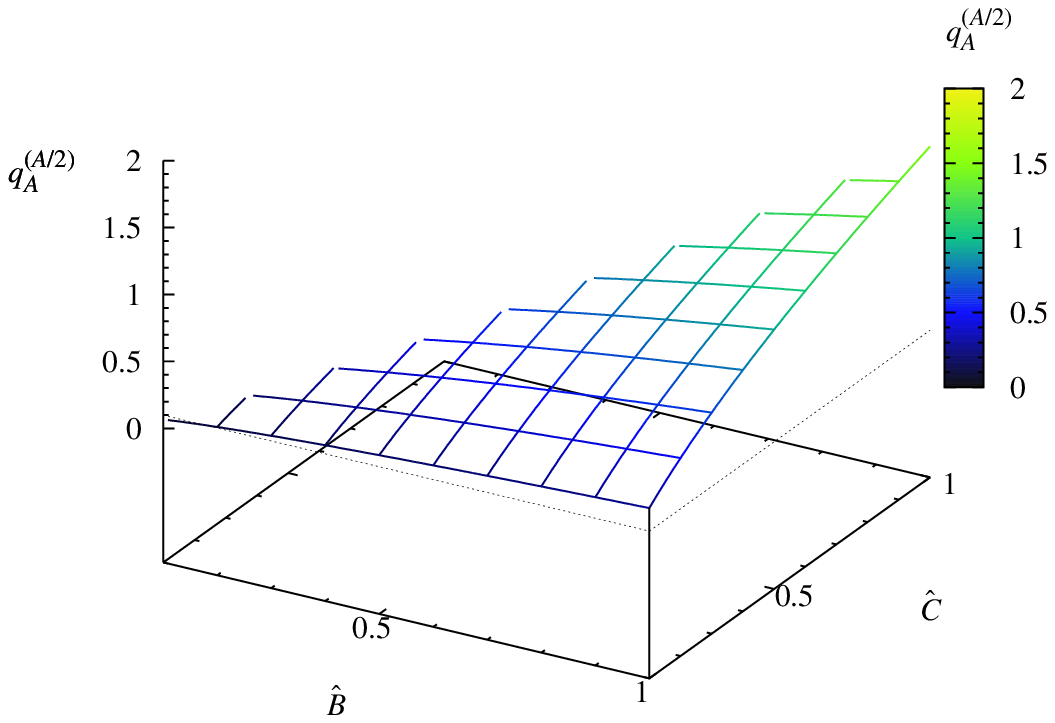}
	\caption{\label{fig:q}$q_A^{(A/2)}$ as a function of $\hat{B}$ and $\hat{C}$ for $\hat{B},\hat{C}\in[0.1,1]$ with $\hat{B} > \hat{C}$. Dotted lines represent $q_A^{(A/2)} = 0.1$.}
\end{figure}

We plot $q_A^{(A/2)}$ in Fig.~\ref{fig:q} as a function of $\hat{B}$ and $\hat{C}$ for $\hat{B},\hat{C}\in[0.1,1]$ with $\hat{B} > \hat{C}$. Its range extends from $0.07$ to $1.5$ on this domain, but it drops below $0.1$ only for $\hat{C}$ very close to $0.1$ and $\hat{B}$ below $0.2$; i.e., for extremely thin rod-like particles, which do not appear to be common based on photographs. The range of $q_A^{(B/2)}$ (not shown) on the same domain is between $0.2$ and $7$, and values above $4$ are again restricted to very small $\hat{C}$ and to $\hat{B} < 0.3$. The results are very similar for other choices of $X$, deviations beyond $20\%$ with respect to $X=A$ are only found for small $\hat{C}$ and do not reach beyond $40\%$ even there.

We have left the question which orientation is relevant open so far. In small-scale isotropic turbulence, which is certainly present in the ocean, nonspherical particles have a preferential alignment with certain characteristics of the flow but undergo rotation \citep{voth2017}. This is why we have chosen to simply cover three perpendicular orientations in our analysis, and have found that differences that may arise from changes in the orientation are minor in most of the domain describing shapes. The only relevant exception is small $\hat{C}$ with $\hat{B} \approx 1$. This regime may characterize paint flakes \citep{song2014,bagaev2017}, but the relative difference remains below $40\%$ even there.

Even though the real advection of the particles will become more complicated as a result of the ever-changing orientation and may thus be beyond the scope of the MRG equation (cf. the discussion in the main text about the settling velocity of irregular particles), we have found that changing orientation introduces minor variations in the value of the settling velocity. Together with the absence of order-of-magnitude corrections that may arise from a nonspherical shape (but comparing shapes under the assumption of the same particle density), this gives quantitative support for the applicability of a spherical shape in Eq.~(\ref{eq:reference}) of the main text.

Finally, we briefly comment on the more general Eqs.~(\ref{eq:inertial}) and (\ref{eq:inertialC}), in which effects from nonsphericity arise in the inertial term through added mass. Since corrections in added mass with respect to a sphere are of order $1$ for all common shapes \citep[see][for an overview]{2014359}, we believe that the finding of App.~\ref{sec:APPeffects} about the negligible importance of inertial effects in these equations is not affected by a deviation from sphericity. The Stokes number, which is proportional to the settling velocity and also depends on the coefficient of added mass ($St \sim \tau_p$ with $\tau_p$ given by Eq.~(\ref{eq:vsbeta})), is estimated to be $10^{-3}-10^{-2}$ for spherical particles in Sect.~\ref{subsec:corrections}, so that it will not increase to $1$ due to a nonspherical shape either, hence leaving the approximation (\ref{eq:inertial}) of Eq.~(\ref{eq:MRG}) valid.

\section{Importance of different physical effects in the dynamics}
\label{sec:APPeffects}

We present here the detailed numerical analysis of the relevance of a finite
time of response (Stokes time, $\tau_p$) of the particle to
the fluid forces, the Coriolis force, and scales unresolved by the NEMO velocity field.


We incorporate the first two effects to a single equation,
\begin{equation}
\textbf{v} = \textbf{u} + \textbf{v}_{s} +
\tau_{p}(\beta-1)\left(\frac{D \textbf{u}}{Dt} + 2\mathbf{\Omega} \times \textbf{u} \right) \ ,
\label{eq:inertialC}
\end{equation}
which is identical to Eq. (\ref{eq:inertial}) except for the
addition of the Coriolis force, $2\mathbf{\Omega} \times
\textbf{u}$. $\mathbf{\Omega}$ is Earth's angular velocity
vector. We include the Coriolis force because it can be more
important than the other inertial term, given by the fluid
acceleration $D\textbf{u}/Dt$, in large-scale ocean flows
$\textbf{u}$ \citep{haller2008inertial,monroy2017modeling}.

The effect of unresolved scales will also be estimated by
keeping the original NEMO velocity field $\textbf{u}$ but
modifying the equation of motion, Eq. (\ref{eq:reference}), by
adding a stochastic noise term:
\begin{equation}
\textbf{v} =\textbf{u} + \textbf{v}_{s} + \textbf{W} \ .
\label{eq:noise}
\end{equation}
$\textbf{W}(t)=(\sqrt{2 D_h}\xi_x(t), \sqrt{2 D_h}\xi_y(t),
\sqrt{2 D_v}\xi_z(t))$, where $\boldsymbol{\xi}(t)$ is a vector Gaussian white
noise process (independent for each particle) of zero mean and with
correlations given by $\langle \xi_{i}(t_{1}) \xi_{j}(t_{2})
\rangle = \delta_{ij} \delta (t_{1}-t_{2})\ ,\ i,j=x,y,z$.
Thus, the horizontal and vertical intensities of this term are
given by $D_h$ by $D_v$, respectively.

The statistical properties are chosen to be
similar to the ones expected for oceanic motions below the
scales resolved by the numerical model
\citep{monroy2017modeling,kaandorp2020closing}. To do so, we
use for $D_h$ Okubo's empirical formulation
\citep{okubo1971oceanic} that parameterizes the effective
horizontal eddy-diffusion below a spatial scale $\ell$ as
$D_{h}(\ell)=2.055 \times 10^{-4} \ell^{1.55} m^{2}/s$. Taking
for $\ell$ the horizontal resolution of our numerical model
(1/12 degrees) we obtain $D_{h}=7.25~m^{2}/s$. Since the Okubo formula is an empirical fit to surface motions, and effective horizontal diffusivity should be weaker below the thermocline, our results provide an upper bound for the error associated with unresolved scales of fluid motion. For the vertical
diffusivity we take $D_{v}=10^{-5} m^{2}/s$.

In order to compare the different equations of motion, we
release a large number $N=78803$ of particles on the
whole Mediterranean at $1~m$ depth on 8 January 2000. We
associate a $\beta$ parameter to each particle by selecting it
from a random uniform distribution in the range $\beta \in
[0.8,1)$, and once it is selected it remains fixed at all times
for the corresponding particle. High values of $\beta$ (close
to one) correspond to more buoyant plastic particles whereas
low values of $\beta$ correspond to high settling velocities.
The corresponding range of velocities is $v_{s} \in [1.776
\times 10^{-3},0)~m/s$. We integrate the particle trajectories
using Eq. (\ref{eq:reference}) and also, from the same initial
conditions, using the corrected dynamics in Eq.
(\ref{eq:inertialC}) or (\ref{eq:noise}), all with the same
NEMO velocity field $\textbf{u}$. The horizontal and vertical
distances between particles released from the same point but
evolved with different equations are compared by calculating
the following averages over particles:
\begin{equation}
d_{h}^I(t)= \frac{1}{N}  \displaystyle\sum_{k=1}^N |\textbf{r}_{k}^{0}(t) - \textbf{r}_{k}^{I}(t)| \  \ , \
d_{v}^I(t)= \frac{1}{N}  \displaystyle\sum_{k=1}^N |z_{k}^{0}(t) - z_{k}^{I}(t)| \ .
\end{equation}
$\textbf{r}_{k}=(x_k,y_k)$ is the horizontal position of
particle $k$, $z_k$ is its vertical position, and the
superindices $0$ and $I$ indicate that the particle trajectory
has been integrated by using the reference equation, Eq.
(\ref{eq:reference}) or the one containing inertial
corrections, Eq. (\ref{eq:inertialC}). The quantities
$d_{h}^W(t)$ and $d_{v}^W(t)$, comparing Eq.
(\ref{eq:reference}) with the dynamics (\ref{eq:noise})
modeling small-scale flow effects, are defined analogously.

\begin{figure*}[t]
\includegraphics[width=16cm]{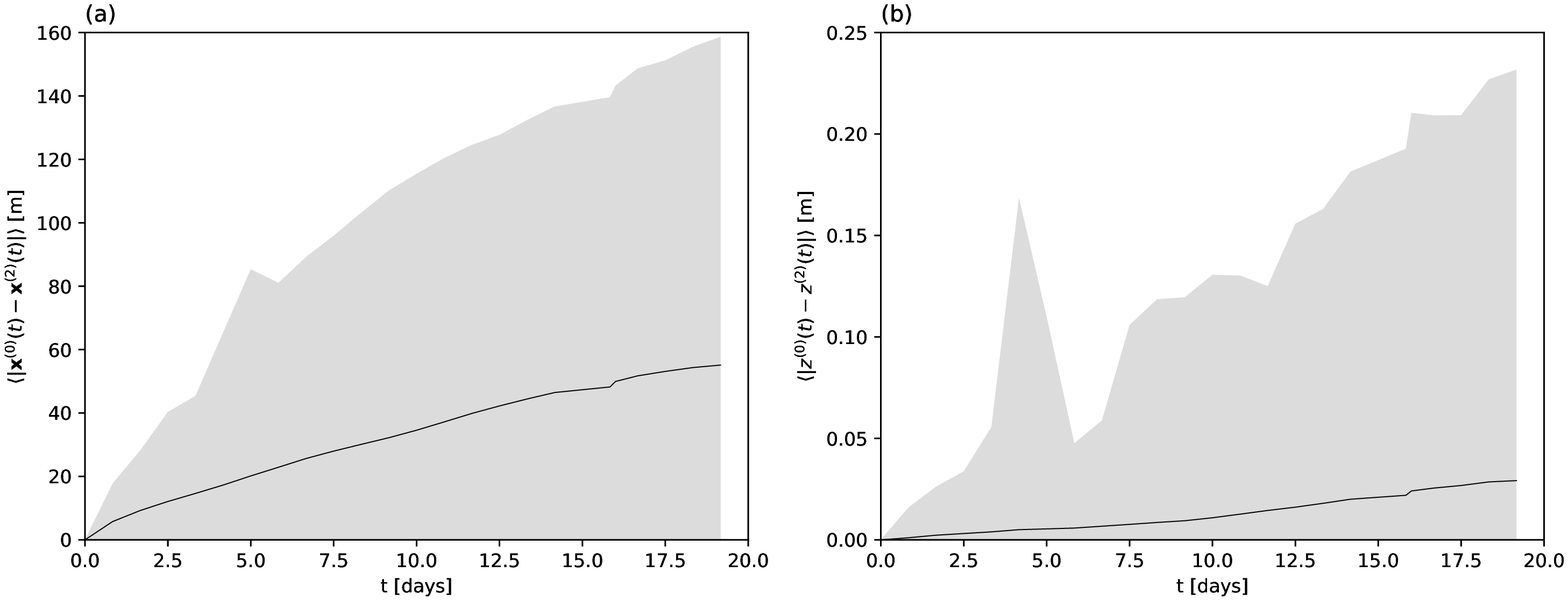}
\caption{\label{Fig:rh02}Solid line indicates the average horizontal distance
$d_{h}^I$ (a) and the average vertical distance $d_{v}^I$ (b)
of particles released from the same initial location but integrated with
equations (\ref{eq:reference}) and (\ref{eq:inertialC}) in the NEMO velocity field.
Shaded region indicates the range of the distances among the individual pairs of particles.}
\end{figure*}

\begin{figure*}[t]
\includegraphics[width=16cm]{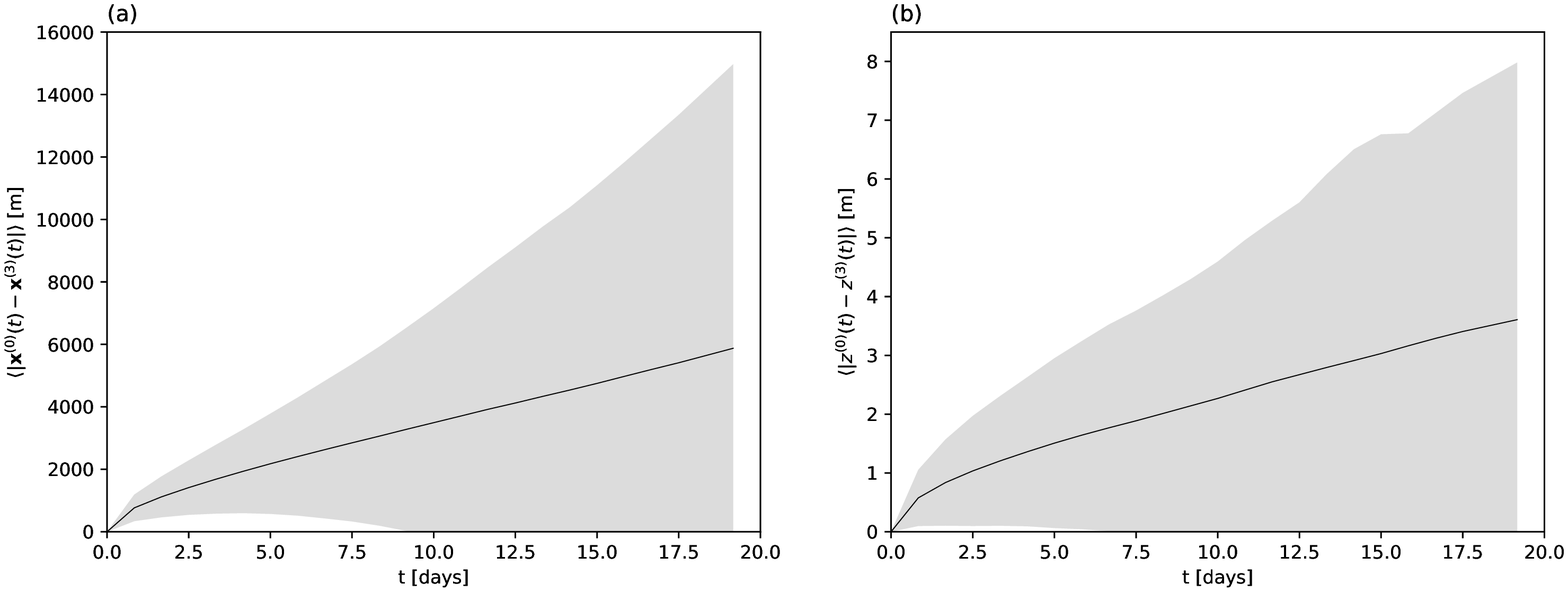}
\caption{\label{Fig:rh03}Same as Fig.~\ref{Fig:rh02} for the comparison of equations (\ref{eq:reference}) and (\ref{eq:noise}).}
\end{figure*}

In Fig. \ref{Fig:rh02}, we display the average distances
$d_{h}^I(t)$ and $d_{v}^I(t)$ as a function of time,
characterizing the corrections by inertial terms to the simple
dynamics of Eq. (\ref{eq:reference}). Analogously, Fig.
\ref{Fig:rh03} displays the average distances $d_{h}^W(t)$ and
$d_{v}^W(t)$ as a function of time, characterizing the
estimated corrections
arising from small scales unresolved by the NEMO velocity field.
The effect induced by the
inertial terms is very small and clearly negligible. The impact
of $\textbf{W}$, and thus of small unresolved scales is larger.

To evaluate the different effects more quantitatively, we
summarize in Table \ref{table:distances}, considering particles
separately in different density ranges (given by the ranges in
$\beta$ and associated $v_s$), the average horizontal and
vertical pairwise particle distances, calculated after
integrating the different dynamics for $10~days$.
To fully appreciate the importance of these numbers, in the two
final columns we compute the horizontal and vertical average of
the total distance $r$ traveled by the particles (using Eq.
(\ref{eq:reference})) during the same interval of time. While
the influence of the inertial terms is completely irrelevant
both in a relative and an absolute sense for any realistic
application, more care has to be taken with regards to the
unresolved scales.
Although the vertical error associated with the latter remains
small, its relative importance hugely increases with decreasing
settling velocity. (Indeed, it would tend to infinity for
approaching neutral buoyancy.) At the same time, the relative
horizontal error is the biggest for the fastest-sinking
particles and is well above $10\%$ for them.

\begin{table*}[t]
\caption{\label{table:distances}Average horizontal and vertical pairwise particle distances $d$ and single-particle displacements $r$ after an integration time of $10$ days. See text. $d/r$ is indicated by percentages in parentheses.}
\begin{tabular}{|l|l||rr|rr||rr|rr||r|r|}
\tophline
$\beta$  & $v_{s}$ ($10^{-3}$ m/s) & \multicolumn{2}{c|}{$d_{h}^I$ (m)} & \multicolumn{2}{c||}{$d_{v}^I$ (m)} & \multicolumn{2}{c|}{$d_{h}^W$ (m)} & \multicolumn{2}{c||}{$d_{v}^W$ (m)} & $r_{h}$ (m) & $r_{v}$ (m) \\
\middlehline
$[0.8,0.85)$ & $[1.78,1.25)$ & 60 & (0.26\%)            & 0.016 & (0.001\%)             & 2675 & (11.8\%)          & 2.1 & (0.16\%)             & 22601          & 1291 \\
$[0.85,0.9)$ & $[1.25,0.79)$ & 42 & (0.14\%)            & 0.010 & (0.001\%)             & 2831 & (9.7\%)           & 2.0 & (0.24\%)             & 29278          & 870 \\
$[0.9,0.95)$ & $[0.79,0.37)$ & 31 & (0.08\%)            & 0.009 & (0.002\%)             & 3313 & (8.0\%)           & 2.1 & (0.43\%)             & 41587          & 493 \\
$[0.95,1)$   & $[0.37,0)$    & 16 & (0.03\%)            & 0.008 & (0.005\%)             & 4668 & (8.0\%)           & 2.7 & (1.79\%)             & 58320          & 151 \\
\bottomhline
\end{tabular}
\end{table*}

\begin{figure*}[t]
\includegraphics[width=16cm]{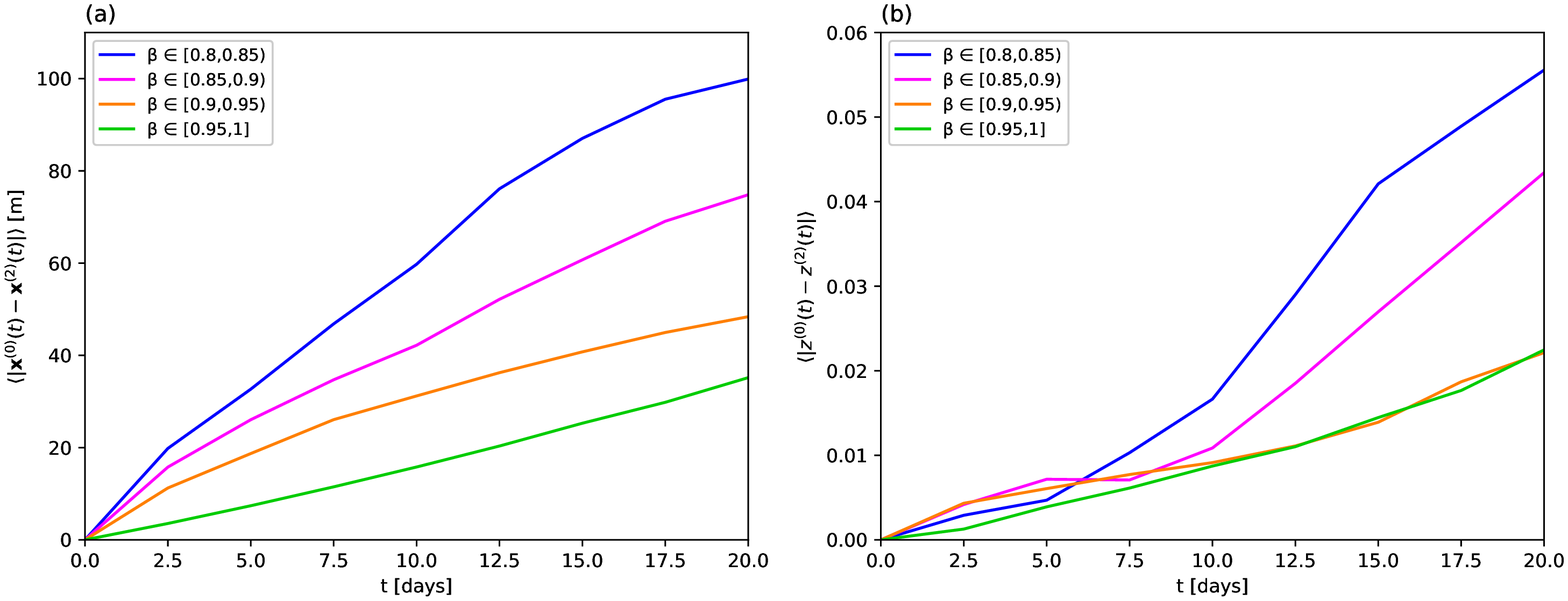}
\caption{\label{Fig:avgbeta1}Average horizontal distance
$d_{h}^I$ (a) and average vertical distance $d_{v}^I$ (b)
of particles released from the same initial location but integrated with
equations (\ref{eq:reference}) and (\ref{eq:inertialC}) in the NEMO velocity field.
The different lines are obtained from
particles from different ranges of densities characterized by the indicated ranges in $\beta$.}
\end{figure*}

\begin{figure*}[t]
\includegraphics[width=16cm]{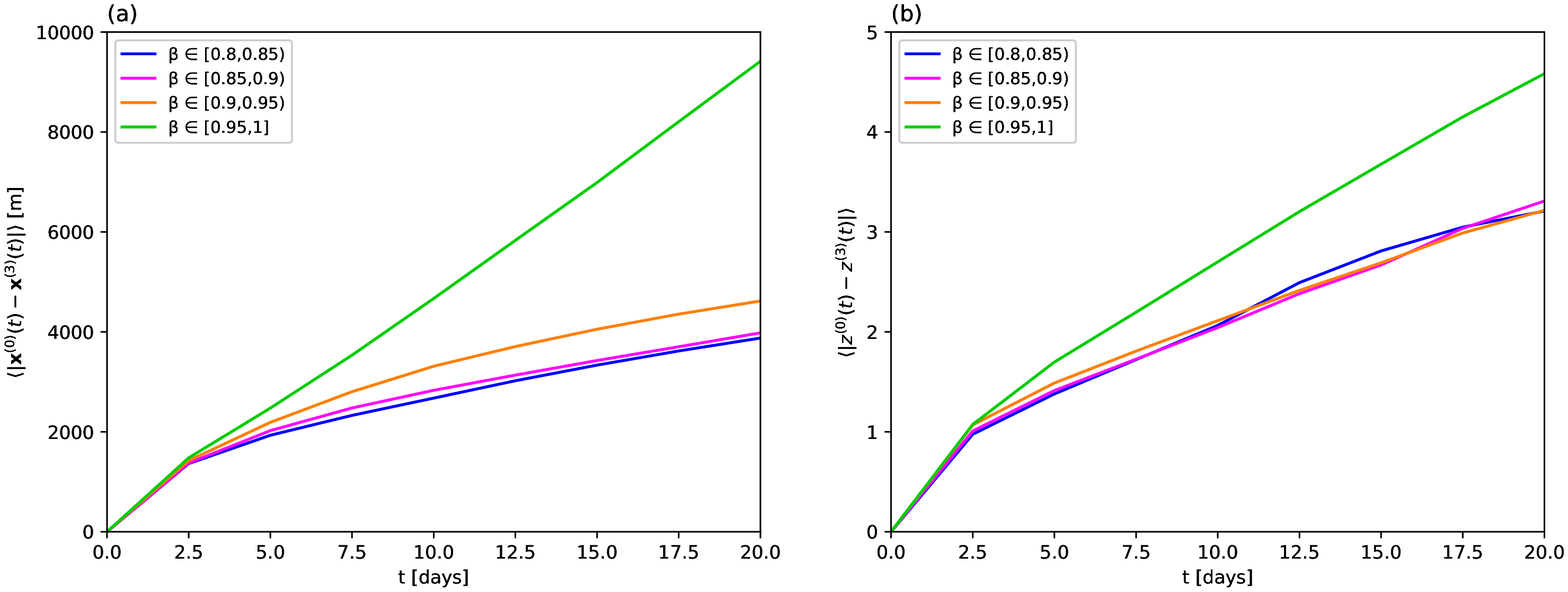}
\caption{\label{Fig:avgbeta2}Same as Fig.~\ref{Fig:avgbeta1} for the comparison of equations (\ref{eq:reference}) and (\ref{eq:noise}).}
\end{figure*}

We can also observe the time evolution of the $d$ entries of this
table in Figs. \ref{Fig:avgbeta1} and \ref{Fig:avgbeta2}.
The overall effect of the unresolved scales is confirmed to be
much larger than that of the inertial terms, and the differences
between particles with different densities (ranges of $\beta$)
are less noticeable (except for the smallest densities
considered, i.e. $\beta\approx 1$).

\section{The effect of bathymetry on vertical dispersion}
\label{sec:APPbathymetry}

We investigate here if the finite and spatially varying depth
of the basin (the bathymetry) could affect the conclusions in
Sect. \ref{subsec:transient} about vertical dispersion. The
falling particles released from different locations reach the
seafloor at different times, and then computing some statistics
over these particles involves
an increasingly narrow set of particles.
Note that the bottom boundary (i.e., the seafloor) extends from
the surface (close to the coast) to the deepest point of the
basin, it is thus relevant at any time during the simulation.
We intend to exclude three different effects arising from the
continual removal of particles: (i) distortion of the shape of
the distribution close to the boundary, (ii) poor quality of
the statistics when many particles have already been lost,
(iii) decrease in the geographical area sampled by the
particles.

\begin{figure}[t]
\includegraphics[width=8.3cm]{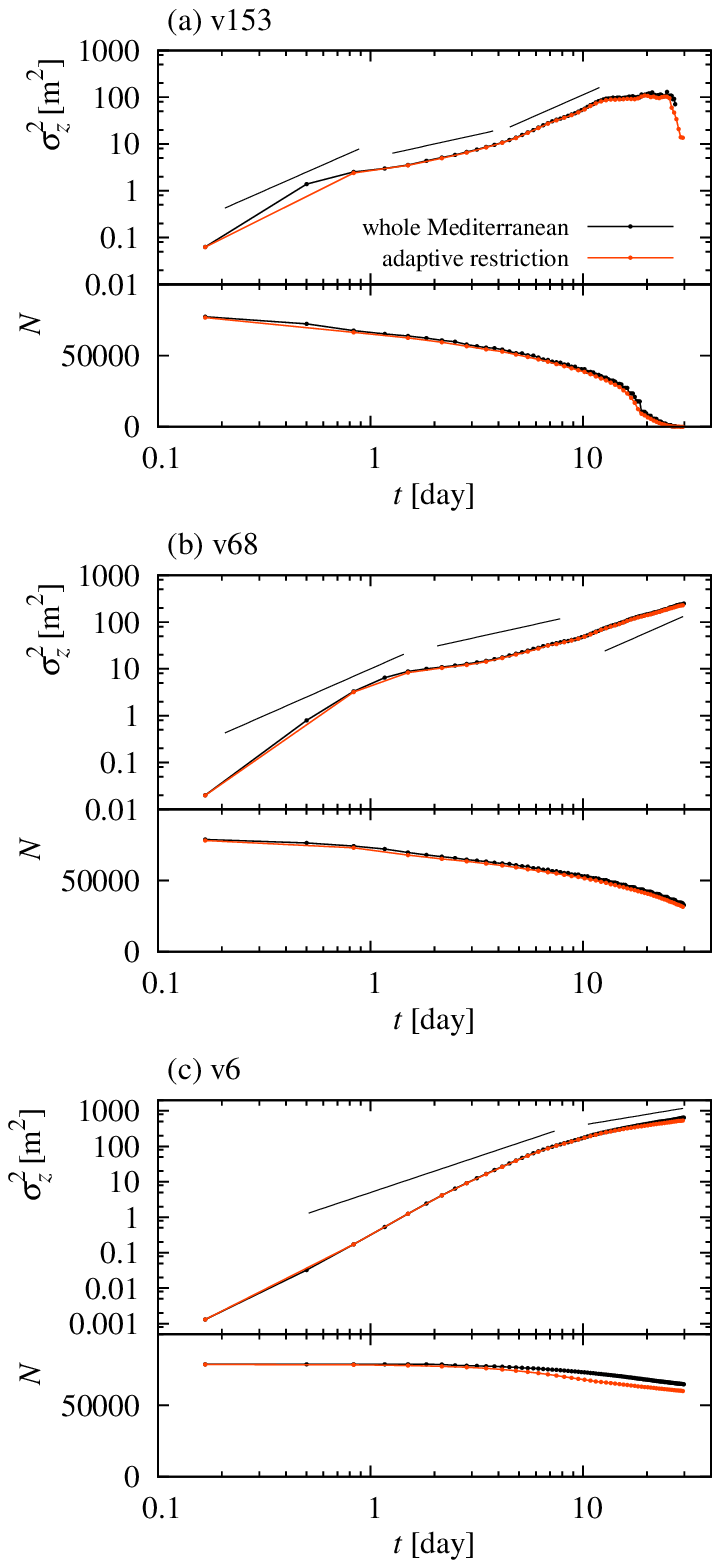}
\caption{\label{Fig:variancetime_withadaptive}Variance of depth reached by all sinking particles (in black) and an
adaptively restricted subset of them (in yellow) as a function of time. See text for details.
Straight lines represent power laws for reference, with exponents $1$ and $2$.
The number $N$ of particles considered for the computation of the variance is also shown.}
\end{figure}

We start with effect (i) by comparing, in
Fig.~\ref{Fig:variancetime_withadaptive}, the variance
presented in the main text (Fig.~\ref{Fig:variancetime}),
computed over all sinking (but not yet sedimented) particles,
and that computed over a restricted set of particles. This
restricted set contains only those particles at the positions
of which the bathymetry $Z$ satisfies $Z > \langle z_i \rangle
- 3 \sigma_z$, where the average $\langle z_i \rangle$ and the
standard deviation $\sigma_z$ are taken with respect to the
original (unrestricted) set of particles. This restriction is
adaptive and ensures that the seafloor is sufficiently far for
its effect on the particle distribution to be negligible at the
positions of all particles kept for the computation.
Fig.~\ref{Fig:variancetime_withadaptive} shows that the
difference in the results between the full set and the
restricted one is negligible for all three settling velocities
considered, except perhaps for the drop at the very end of the
time evolution, after the constant section, in
Fig.~\ref{Fig:variancetime_withadaptive}a. This drop is,
however, very short compared to the bulk of the sinking process
and thus have minor importance. Furthermore, the distribution
of particles so close to the bottom (cf.
Fig.~\ref{Fig:variancedepth}) should anyway be strongly
influenced by resuspension and remixing by bottom currents
\citep{kane2020seafloor}.
For the rest of the time evolution, we can be confident that a
possible distortion of the particle distribution induced by the
boundaries has no impact on the variance curves. Such
distortion may be present, but the number of particles close to
the seafloor is very small at any given time instant (see the
small difference between the numbers $N$ of particles
considered for the two kinds of computation in
Fig.~\ref{Fig:variancetime_withadaptive}) so that they have a
negligible contribution on the originally computed variance.
This is a result of the relatively weak dispersion; artifacts
might be found for broader distributions, possibly including a
hypothetical continuation of the v6 simulation. The time evolution of $N$ in
Fig.~\ref{Fig:variancetime_withadaptive} also assures us that
poor-quality statistics, effect (ii), does not arise until the
final drop discussed in the previous paragraph. Except for that
very final section, the variance is estimated from a
sufficiently large number of particles to keep the relative
error of the sample variance (with respect to the population
variance) very low. This is so because, under standard
assumptions, the ratio between the sample and the population
variance should be close to a chi-square random variable with
$N-1$ degrees of freedom, divided by $N-1$ \citep{douillet2009}.

The time evolution of $N$ also suggests that effect (iii) is
avoided as well:
in Fig.~\ref{Fig:variancetime_withadaptive}, there is no sudden
drop in the number of particles during the simulations that
could result in the changes in the slope of the curves of
variance versus time. To further support this conclusion, we
compute the time evolution of the variance over the particles
initialized in a subregion of the whole Mediterranean, see in
Fig.~\ref{Fig:08HorizontalVm}. In particular, we choose the box
of longitudes $[5,6.9]$ degrees, and latitudes $[37.5,42]$
degrees, corresponding to the Sea of Sardinia, where the
bathymetry is deep enough to prevent particles from reaching
the seafloor (except for the very last few time steps of the
v153 configuration), so that the horizontal area sampled by the
particles approximately remains constant (remember that
horizontal displacements are small compared to geographical
features).

\begin{figure}[t]
\includegraphics[width=8.3cm]{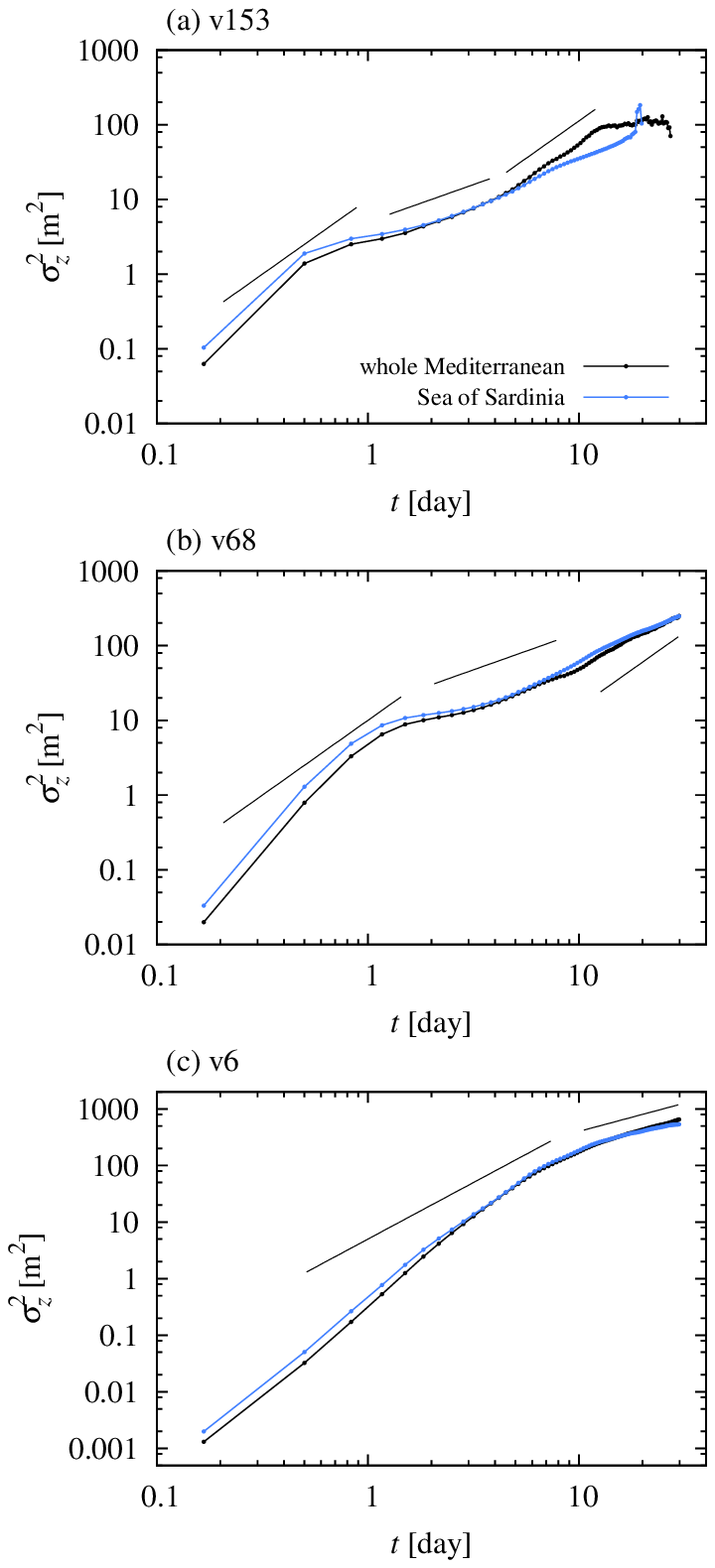}
\caption{\label{Fig:variancetime_withSardinia}Variance of depth reached by all sinking particles (in black)
and those initialized in the Sea of Sardinia (in blue; longitudes in
$[5,6.9]$ degrees, and latitudes in $[37.5,42]$ degrees)
as a function of time. See text for details. Straight lines
represent power laws for reference, with exponents $1$ and $2$.}
\end{figure}

According to Fig.~\ref{Fig:variancetime_withSardinia}, the
character of the dispersion in the Sea of Sardinia is nearly
identical to that in the whole Mediterranean, except after the
crossover from normal diffusive to ballistic dispersion in the
v153 case (Fig.~\ref{Fig:variancetime_withSardinia}a). The
smaller variance in the velocity should naturally lead to a
later crossover to the long-time behavior in the Sea of
Sardinia (see Section~\ref{subsec:transient}), it is
nevertheless doubtful that the crossover should fall outside
the simulation time. As the crossover is not observed, one
might speculate that horizontal mixing might just get strong
enough to suppress ballistic dispersion, similarly to the v68
case (see the discussion of Fig.~\ref{Fig:variancetime}), for
which the time evolution of the variance is very similar in the
Sea of Sardinia and the whole Mediterranean
(Fig.~\ref{Fig:variancetime_withSardinia}b).
Fig.~\ref{Fig:08HorizontalVm} suggests that the characteristic
patch size is smaller in the western basin of the
Mediterranean, including the Sea of Sardinia, than in the
(considerably larger) eastern one, which makes inter-patch
mixing easier. We conclude that the only substantial difference
between the dispersion in the Sea of Sardinia and the whole
Mediterranean may
originate from the smaller extension and some special
characteristics of the former, and the decrease in the area
sampled in the whole-Mediterranean simulation presumably has
not effect on the results.

Based on these analyses, we believe that the findings of
Section~\ref{subsec:transient} are unaffected by the boundary
and thus have general relevance for oceanic dispersion.

\noappendix       







\authorcontribution{
All authors designed research. RF performed the simulations. RF and
GD analyzed data. RF, GD, EHG and CL prepared the first draft, and all authors reviewed
and edited the manuscript.}


\competinginterests{The authors declare that they have no conflict of interest.} 


\begin{acknowledgements}
This work is part of the ``Tracking of Plastics in Our Seas''
(TOPIOS) project, funded by the European Research Council under
the European Union's Horizon 2020 research and innovation
program (grant agreement no. 715386). We acknowledge publication
fee support from the CSIC Open Access Publication Support Initiative
through its Unit of Information Resources for Research (URICI). R.F., G.D., E.H-G and
C.L acknowledge financial support from the Spanish State
Research Agency through the Mar\'{\i}a de Maeztu Program for
Units of Excellence in R\&D (MDM-2017-0711). R.F. also
acknowledges the fellowship no. BES-2016-078416 under the FPI
program of MINECO, Spain. G.D. also acknowledges financial
support from the European Social Fund through the fellowship
no. PD/020/2018 under the postdoctoral program of CAIB, Spain,
and from NKFIH, Hungary (grant agreement no. NKFI-124256).
\end{acknowledgements}



%
%
%

\bibliographystyle{copernicus}
\bibliography{Manuscript}

\end{document}